\documentclass[amsmath,amssymb,aps,prd,10pt,floatfix,preprintnumbers,superscriptaddress,nofootinbib,twocolumn]{revtex4-1}
\usepackage{graphicx}
\usepackage{lineno}
\usepackage{nicefrac}
\usepackage[normalem]{ulem}
\usepackage{subfigure}  
\usepackage{multirow}
\usepackage{hyperref}
\usepackage[all]{nowidow}
\usepackage[T1]{fontenc}
\def\ncteq{{nCTEQ}}
\def\ncteqpp{\hbox{nCTEQ++}}
\def\ncteqwz{\hbox{nCTEQ15WZ}}
\def\ncteqhix{{nCTEQ15HIX}}

\def\chidof{\chi^2/N_{dof}}
\def\GeV{\text{GeV}}

\newcommand{\orcid}[1]{\,\href{https://orcid.org/#1}{\includegraphics[width=9pt]{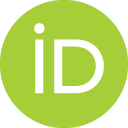}}\,}

\newcommand{\orcidES}{0000-0001-5606-0350} %
\newcommand{\orcidTJ}{0000-0002-1334-7607} %
\newcommand{\orcidAA}{0000-0002-2077-6557} %
\newcommand{\orcidPD}{0000-0001-7960-7953} %
\newcommand{\orcidOH}{0000-0002-4890-6544} %
\newcommand{\orcidTH}{0000-0002-2729-0015} %
\newcommand{\orcidTK}{0000-0002-7516-8292} %
\newcommand{\orcidMK}{0000-0002-4665-3088} %
\newcommand{\orcidKK}{0000-0003-1412-447X} %
\newcommand{\orcidAK}{0000-0002-4090-0084} %
\newcommand{\orcidJM}{0000-0001-9343-9351} %
\newcommand{\orcidFM}{0000-0002-3888-1697} %
\newcommand{\orcidFO}{0000-0001-6799-2436} %
\newcommand{\orcidIS}{0000-0003-0373-474X} %
\newcommand{\orcidJY}{0000-0001-8366-0968} %
\begin{document}

\setlength\linenumbersep{2pt}

\preprint{FNAL-PUB-20-668} 
\preprint{JLAB-THY-20-3303}
\preprint{SMU-HEP-20-07}
\preprint{MS-TP-20-40}
\preprint{KA-TP-16-2020}%
\preprint{P3H-20-052}%
\preprint{IFJPAN-IV-2020-9}

\def\mit{\affiliation{Massachusetts Institute of Technology, Cambridge, Massachusetts 02139, USA}}
\def\smu{\affiliation{Department of Physics, Southern Methodist University,
    Dallas, TX 75275-0175, U.S.A.}}
\def\jlab{\affiliation{Jefferson Lab, Newport News, VA 23606, U.S.A.}}
\def\kit{\affiliation{Institute for Theoretical Physics, KIT,  D-76131  Karlsruhe, Germany}}
\def\krakow{\affiliation{Institute of Nuclear Physics Polish Academy of Sciences, PL-31342 Krakow, Poland}}
\def\muenster{\affiliation{Institut f{\"u}r Theoretische Physik, Westf{\"a}lische Wilhelms-Universit{\"a}t M{\"u}nster,             \\Wilhelm-Klemm-Stra{\ss}e 9, D-48149 M{\"u}nster, Germany}}
\def\lpsc{\affiliation{Laboratoire de Physique Subatomique et de Cosmologie, Université Grenoble-Alpes, 
    \\CNRS/IN2P3, 53 avenue des Martyrs, 38026 Grenoble, France}}
\def\hampton{\affiliation{Hampton University, Hampton, VA 23668, USA}}
\def\fnal{\affiliation{Fermi National Accelerator Laboratory, Batavia, Illinois 60510, USA}}
\def\iit{\affiliation{Department of Physics, Illinois Institute of Technology, Chicago, Illinois 60616, USA}}

\title{\ncteqhix{} --- Extending nPDF Analyses into the High-\texorpdfstring{\boldmath$x$}{x}, Low-\texorpdfstring{$Q^2$}{Q2}  Region}

\author{E.P.~Segarra\orcid{\orcidES}}
\email{segarrae@mit.edu}
\mit{}

\author{T.~Je\v{z}o\orcid{\orcidTJ}}
\email{tomas.jezo@kit.edu}
\kit{}

\author{A.~Accardi\orcid{\orcidAA}} 
\hampton{} \jlab{} 
\author{P.~Duwent\"aster\orcid{\orcidPD}}
\muenster{}
\author{O.~Hen\orcid{\orcidOH}} 
\mit{}
\author{T.J.~Hobbs\orcid{\orcidTH}}
\smu{} \jlab{} \iit{}
\author{C.~Keppel\orcid{\orcidTK}} 
\jlab{}
\author{M.~Klasen\orcid{\orcidMK}}
\muenster{}
\author{K.~Kova\v{r}\'{i}k\orcid{\orcidKK}} 
\muenster{}
\author{A.~Kusina\orcid{\orcidAK}} 
\krakow{}
\author{J.G.~Morf\'{i}n\orcid{\orcidJM}} 
\fnal{}
\author{K.F.~Muzakka\orcid{\orcidFM}}
\muenster{}
\author{F.I.~Olness\orcid{\orcidFO}} 
\email{olness@smu.edu}
\smu{}
\author{I.~Schienbein\orcid{\orcidIS}} 
\lpsc{}
\author{J.Y.~Yu.\orcid{\orcidJY}}
\lpsc{}

\begin{abstract}
\null \vspace{0.5cm}
We use the \ncteq{} analysis framework to investigate nuclear Parton Distribution Functions (nPDFs) in the region of large $x$ and intermediate-to-low $Q$, with special attention to recent JLab Deep Inelastic Scattering data on nuclear targets. This data lies in a region which is often excluded by $W$ and $Q$ cuts in global nPDF analyses.
As we relax these cuts, we enter a new kinematic region, which introduces new phenomenology. 
In particular, we study the impact of 
i)~target mass corrections, 
ii)~higher twist corrections,
iii)~deuteron corrections, 
and
iv)~the shape of the nuclear PDF parametrization at large-$x$ close to one.
Using the above tools, we produce a new nPDF set (named \ncteqhix{}) which 
yields a good description of the new JLab data
in this challenging kinematic region, and displays reduced uncertainties at large $x$, in particular for up and down quark flavors.
\null \vspace{0.5cm}
\end{abstract}

\date{\today}

\maketitle
\tableofcontents{}

\newpage
\begin{figure}[tb]
\includegraphics[width=0.45\textwidth]{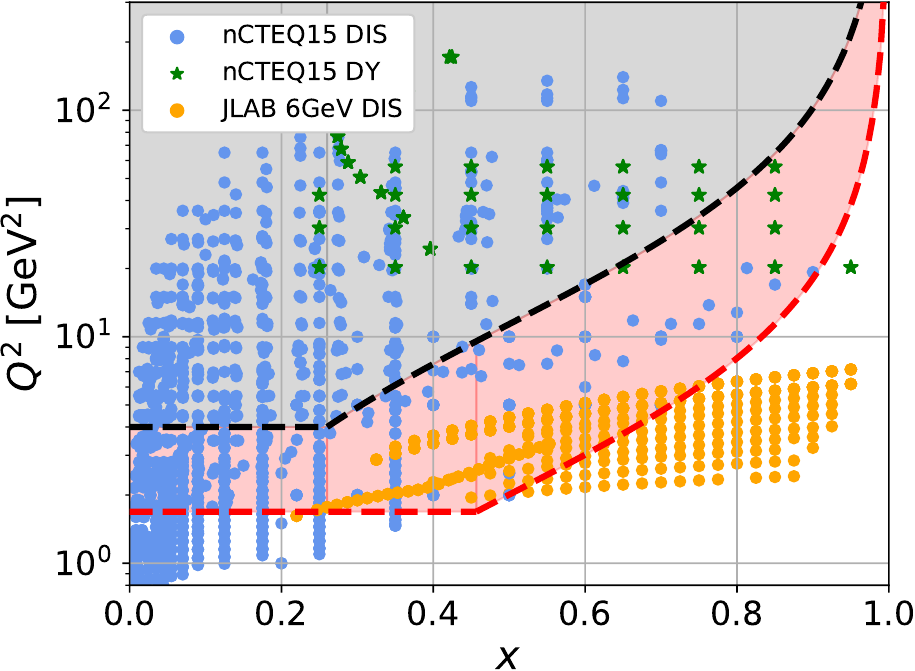}
\caption{We display DIS and DY data entering our analysis
in the $\{x,Q^2\}$ space indicating the relevant kinematic cuts,
where $x$ and $Q^2$ are the usual DIS variables, and 
$Q^2$ for DY  is the di-lepton mass squared. 
The more restrictive cuts of
$Q=2$~GeV and $W=3.5$~GeV (black dashed line)
are the cuts used in the original nCTEQ15 analysis.
In the present work, we will relax the cuts to 
$Q=1.3$~GeV and $W=1.7$~GeV (red dashed line).
This greatly expands the kinematic reach in the high-$x$ region
where much of the new JLab data is located.  
}
\label{fig:kin}
\end{figure}
\begin{figure}[tb]
\includegraphics[width=0.45\textwidth]{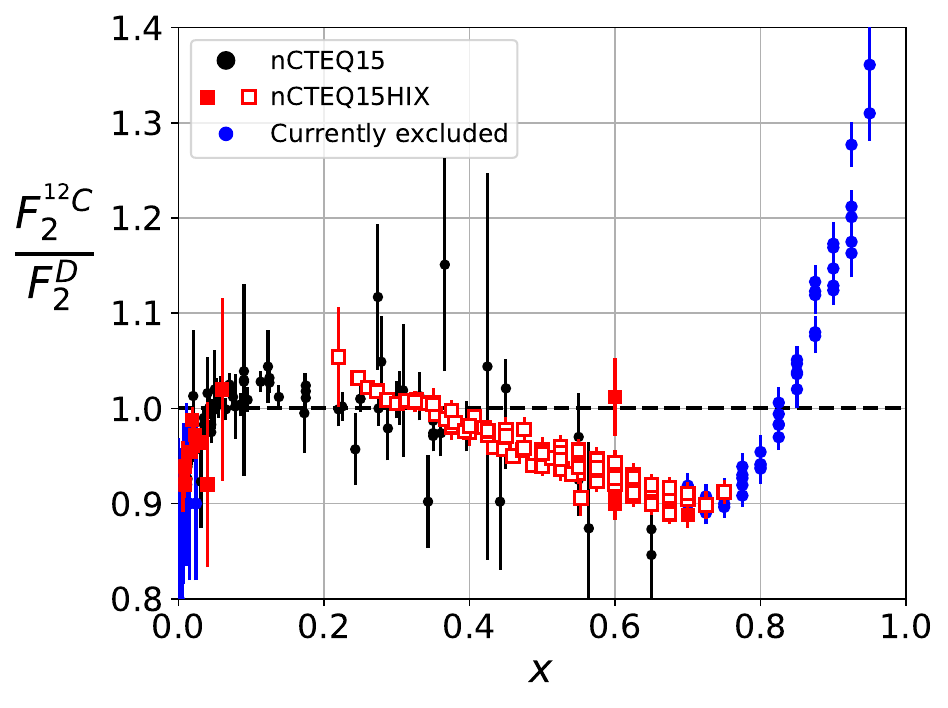} %
\caption{We display the classic $F_2^A/F_2^D$ ratio for carbon illustrating
the nuclear correction factor across the various $x$ regions. 
The  black points indicate the data used in the original nCTEQ15 fit,
and the red points with the solid squares represent the additional data from 
this original set which are now included due to the relaxed $Q$ and $W$ cuts. 
The red open squares are the new  JLab DIS data included in this analysis,
and the blue points are those JLab DIS data which are excluded by the current kinematic cuts.
}
\label{fig:emc}
\end{figure}
\begin{table}[tb]
\vspace{0.5cm}
\centering{}%
\begin{tabular}{|c|c||c|c|c|c|c|}
\hline 
 &  & $W_{cut}$ & $W_{cut}$ & $W_{cut}$ & $W_{cut}$ & $W_{cut}$\tabularnewline
$Q_{cut}^{2}$ & $Q_{cut}$ & {\scriptsize No Cut} & 1.3 & 1.7 & 2.2 & 3.5\tabularnewline
\hline 
\hline 
1.3 & $\sqrt{1.3}$ & 1906 & 1839 & 1697 & 1430 & 1109 \tabularnewline
\hline 
1.69 & 1.3 & 1773 & 1706 & 1564 & 1307 & 1024\tabularnewline
\hline 
2 & $\sqrt{2}$ & 1606 & 1539 & 1402 & 1161 & 943\tabularnewline
\hline 
4 & 2 & 1088 & 1042 & 952 & 817 & 708 \tabularnewline
\hline 
\end{tabular}
\caption{The table shows the number of remaining  
data points after the  $\{Q^2,W\}$ kinematic cuts,
where $x$ and $Q^2$ are the usual DIS variables, and 
$Q^2$ for DY  is the di-lepton mass squared. 
The units of $Q$ and $W$ are both in GeV, and $Q^2$ in GeV${}^2$.
For reference, nCTEQ15 used cuts of $Q=2$~GeV and $W=3.5$~GeV,
while the current \ncteqhix{} set uses cuts of  $Q=1.3$~GeV and $W=1.7$~GeV.
}
\label{tab:datacuts}
\end{table}

\section{Introduction}
\label{sec:intro}
With the EIC and LHeC/FCC on the horizon, science is now entering a new era of precision
in the investigation of hadronic structure enabled by a flood of data from JLab, RHIC and the LHC. 
Describing one of the four fundamental forces of nature, Quantum ChromoDynamics (QCD) --- the 
theory of the strong interaction --- remains deeply complex and enigmatic, although the Parton 
Distribution Function (PDF) framework has proven remarkably successful in describing processes 
with hadronic initial states~\hbox{\cite{Kovarik:2015cma,Hirai:2007sx,deFlorian:2011fp,Eskola:2016oht,Walt:2019slu,AbdulKhalek:2019mzd,AbdulKhalek:2020yuc,Ethier:2020way,Paukkunen:2020rnb,Kusina:2016fxy,Khanpour:2020zyu,Nadolsky:2008zw,Ball:2009mk,Ball:2017nwa,Alekhin:2017olj,Lin:2017snn,Gao:2017yyd,Khalek:2018mdn,Kovarik:2019xvh,Hou:2019efy,Sato:2019yez,Harland-Lang:2014zoa,Thorne:2019mpt,Lin:2020rut,Faura:2020oom,AbdulKhalek:2021gbh}}.

While the study of proton PDFs has grown exceedingly precise, the need to extend this
precision to the nuclear sector, involving fits with explicit nuclear degrees of freedom,
has become more urgent in recent years in order to enhance the accuracy of experimental 
analyses involving nuclear targets.
Progress in studying QCD dynamics within nuclei has been demonstrated across a number of 
recent nuclear PDF (nPDF) 
analyses~\hbox{\cite{Kovarik:2015cma,Hirai:2007sx,deFlorian:2011fp,Eskola:2016oht,Walt:2019slu,AbdulKhalek:2019mzd,AbdulKhalek:2020yuc,Ethier:2020way,Paukkunen:2020rnb,Kusina:2016fxy,Khanpour:2020zyu}}.
A significant challenge in the determination of nPDFs has been the acquisition of empirical
data from a sufficiently wide variety of experiments as to provide complementary
constraints, and, {\it e.g.}, specify the $A$ dependence of the resulting nPDFs. For this
reason, there is a continual need for new data sets to broaden global analyses.  In the 
present work, we build upon the recent nCTEQ15 analysis by including recent JLab data 
covering an expanded kinematic range. As we shall demonstrate, this data has the potential
to furnish an improved understanding of hadronic and nuclear structure and interactions, and,
in turn, new insights into QCD.

\subsection{JLab Kinematic Reach}
\label{sec:introJLab}

The recent facility upgrades of the 
Continuous Electron Beam Accelerator Facility
at the Thomas Jefferson National Accelerator Facility (JLab)
have enabled the measurement of high precision electron-nucleus scattering events 
in an extended kinematic regime. 
In particular, the JLab experiments provide a wealth 
of data in the relatively unexplored kinematic region of 
large Bjorken $x$ and intermediate to low photon virtuality $Q^2$. This mostly unexplored kinematic region is often referred to as the ``transition'' region from resonance dominated production to deep-inelastic scattering, and is of considerable interest to both the charged lepton and neutrino scattering communities.

In Fig.~\ref{fig:kin} we display a collection of DIS and DY data
from selected experiments along with  
recent data from the JLab DIS experiments. 
We have also indicated typical cuts in $Q^2$ and 
$W^2=Q^2(1-x)/x + M_N^2$, with $M_N$ the nucleon mass, which are often implemented 
in many global analyses.
The objective of these $\{W,Q\}$ cuts is to remove
those data which might have significant non-perturbative 
or higher twist contributions; unfortunately, these  cuts 
exclude a large subset of the data, as indicated in Table~\ref{tab:datacuts}.

In this investigation we will relax the kinematic cuts 
to study whether we can describe the broader subset of the JLab data 
with nuclear PDFs using the \ncteq{} global analysis framework~\cite{Kovarik:2015cma}.
As we expand this analysis into new kinematic 
regions, we will analyze   whether we might be sensitive 
to new effects including 
i)~target mass corrections, 
ii)~higher twist corrections,
iii)~deuteron structure function modifications, 
and
iv)~large-$x$ nuclear corrections to allow the nuclear Bjorken variable $x_A>1$.
We will address each of these issues in turn,
and then use the optimal combination to extract 
a set of nPDFs which extends into this new kinematic region. 

We note here a recent study~\cite{Paukkunen:2020rnb}
where PDF reweighting methods were used to investigate compatibility
of the current nuclear PDFs with the CLAS data.

\subsection{The EMC Effect}
\label{sec:emc}

Extending the nPDFs into this new kinematic region can also provide new information on  $x$-dependent nuclear effects observed in many DIS experiments.
Starting in the early 1980's, the European Muon Collaboration (EMC)~\cite{Aubert:1983xm} found that in the DIS kinematic region the per-nucleon structure functions $F_{2}$ for iron and deuterium were not only different, but that this difference also changed as a function of $x$. 

This evidence for nuclear effects in DIS charged-lepton nucleus scattering can be summarized as in Fig.~\ref{fig:emc}, which displays the $F_{2}^{C}/F_{2}^{D}$ per-nucleon structure function ratio as a function of $x$; 
this behavior is consistent with measurements  by both 
the SLAC $e A$~\cite{Gomez:1993ri,Dasu:1993vk,Bodek:1983qn} and 
the BCDMS $\mu A$~\cite{Bari:1985ga,Benvenuti:1987az} experiments.
The observed behavior of the ratio $R{=}F_{2}^{A} / F_{2}^{D}$,
can be divided into four $x$ regions: 

\begin{itemize}
\item the shadowing region: $R \leq1$ for $x \lesssim  0.1$,
\item the anti-shadowing region: $R \geq 1$ for  \hbox{$0.1 {\lesssim}\, x \, {\lesssim} 0.3$},
\item the EMC-effect region: $R \leq 1$ for $0.3 \lesssim x \lesssim 0.7$,
\item and the Fermi motion region:   for $x \gtrsim  0.7$.
\end{itemize}
For a review of the data and theoretical interpretations see Ref.~\cite{Malace:2014uea}.

The current study concentrates on the higher $x$ region so the effects of shadowing and anti-shadowing are 
not of direct impact for this specific study. 
However, the modifications at medium-to-higher-$x$ in the so-called ``EMC-effect region'' 
will  be especially important for these new JLab data sets. 
Since its discovery, there is still no universally accepted explanation for the EMC effect. Recent studies have observed a correlation between the magnitude of the EMC effect and the relative amount of short-range correlated (SRC) nucleon pairs in different nuclei,
suggesting that the EMC effect is driven by the modification of nucleons in SRC pairs~\hbox{\cite{Segarra:2019gbp,Hen:2016kwk,Hen:2013oha,Weinstein:2010rt,Schmookler:2019nvf}}. Forthcoming experiments at JLab will explore the relationship between the offshellness of nucleons and the modification of the structure function~\cite{band-proposal,BANDandLAD}.
Additionally, it has  been demonstrated~\cite{Arrington:2003nt} that the EMC effect persists
to lower values of $W$  extending into the resonance region. 
This is an intriguing result that the nuclear structure functions in the resonance region exhibit similar behavior as in the DIS region.  
In Sec.~\ref{sec:noWcut} we will explore implications of removing the $W$ cut on the EMC effect.

\subsection{Outline}
\label{sec:outline}

In Section~2 we present an overview of the various corrections we will apply 
to better describe the large $x$ region. 
In Section~3 we introduce the new JLab  data used in this study
and summarize the various fits we will investigate. 
In Section~4 we show the results of the fits including the comparison between data and theory, 
and present the impact on the PDFs. 
In Section~5 we explore the implications of extending the nPDFs into the low $W$ kinematic region.
In Section~6 we provide a comparison of the new \ncteqhix{} nPDFs with other results from the literature. 
In Section~7 we summarize our conclusions.
Additionally, 
in Appendix~\ref{sec:datatabs} we tabulate the data sets used in the fit and provide the impact of the kinematic cuts.

\section{Structure Function Modifications} \label{sec:corr}

The goal of this study is to extend our global nPDF fit into 
new kinematic regions at large $x$ and low $Q$. 
Many global analyses impose stringent cuts in both $Q$ and $W$.
For example, the nCTEQ15 fit was performed with $Q{>}2~{\rm GeV}$
and 
$W{>}3.5~{\rm GeV}$.
The motivation for these cuts is that the $Q$ cut largely eliminates 
a variety of nonperturbative and/or power-suppressed corrections with the
potential to complicate the extraction of leading-twist PDFs.
As $Q$ decreases,  $\alpha_s(Q^2)$ becomes large, and our perturbation expansion breaks down. 
Correspondingly, the $W$ cut removes events in both the low $Q^2$ and large $x$ region
where contributions from non-factorizable higher-twist terms become large. 

Imposing stringent cuts allows us to avoid kinematic regions that may be difficult to compute.
However, the trade-off is that this can significantly reduce the data set excluding kinematic regions important for both the charged lepton and neutrino communities. 
In Table~\ref{tab:datacuts} we display the total number of data points which satisfy various kinematic cuts. 
For example, with the loosest set of cuts 
$\{Q{>}1.14~\GeV, W{>}1.3~\GeV\}$, 
the number of data points is 1839; 
this is in contrast to the very conservative cuts used in the nCTEQ15 fit, 
$\{Q{>}2~\GeV, W{>}3.5~\GeV\}$,  
which reduces the number to 708---less than half.

Clearly, it is advantageous to reduce the $\{Q,W\}$ cuts
as much as possible, but this region is 
challenging to compute. 
For example, if we simply take the nCTEQ15 fit and extend this into the low   $\{Q,W\}$  region
without accommodating any additional phenomenological effects, 
the quality of the resulting 
theory to data comparison is  no longer acceptable. 

As we mentioned in the introduction, there are a variety of possible effects that may enter our analysis
in this region, and we will examine them systematically to study the impact of each one to 
obtain the best description of the data. 

In the remainder of this section, we will review the different effects, and 
outline how they are included in the global fit.

\subsection{Target Mass Corrections}
\label{sec:tmc}

\begin{figure}[tb]
\includegraphics[width=0.45\textwidth]{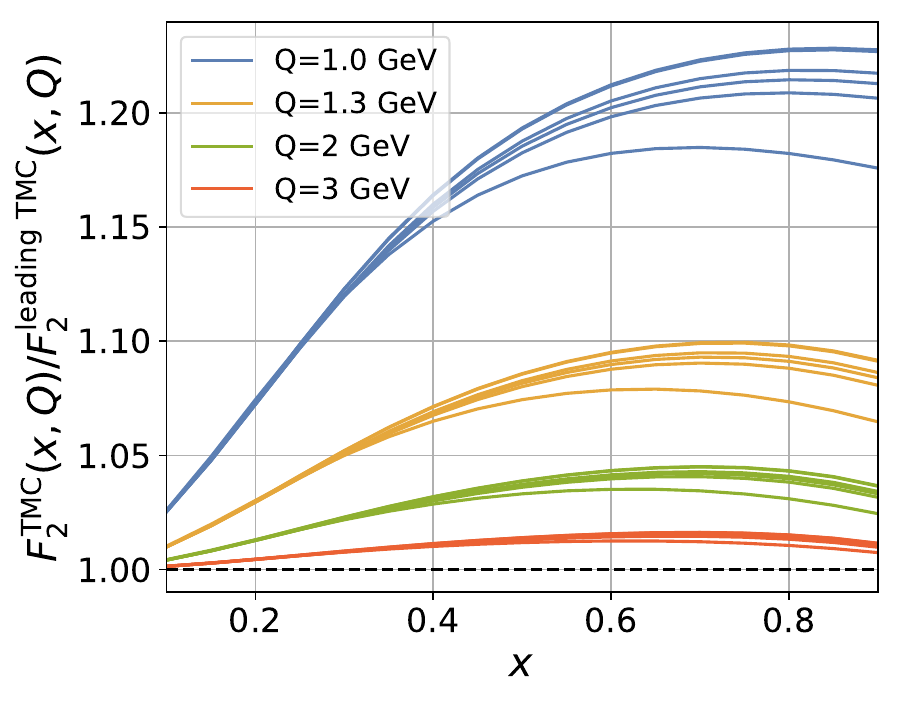}
\caption{We display the ratio 
$F_2^{TMC}(x,Q)/F_2^{\mathrm{{leading}\atop{TMC}}}(x,Q)$
for  $Q=\{1,1.3,2,3\}$ GeV (blue, yellow, green, red) 
for neutral current DIS; the lowest $Q$ values yield the largest corrections. 
Each colored band contains results for \{D, C, Al, Fe, Au, Pb\} nuclei,
where D is the lowest and Pb is the highest within each band. 
}
\label{fig:tmc}
\end{figure}

We begin by examining  the  operator product expansion (OPE)
to determine what possible corrections we might encounter as we extend the 
kinematic region to high-$x$ and low $Q$.
The expressions for the structure functions can be derived in the OPE,
and a detailed review is provided in Ref.~\cite{Schienbein:2007gr}.
Using this formalism, the leading, twist-2, structure function expressions will have corrections 
in powers of $(M/Q)$ where $M$ is the mass of the 
proton:\footnote{%
See Eq.\ (23) of Ref.~\cite{Schienbein:2007gr} for details.
We have implemented an alternative definition derived in the light-cone frame~\cite{Aivazis:1993kh,Aivazis:1993pi}
which coincides with Eq.~(\ref{eq:F2TMC}) to ${\lesssim} 1\%$.
} %
\begin{equation}
    F_2^\mathrm{TMC}(x,Q) = \frac{x^2}{\xi^2 r^3} F_2^{(0)}(\xi,Q) + \ldots
\label{eq:F2TMC}
\end{equation}
Here, $\xi{=}2x/(1{+}r)$ is the Nachtmann variable with $r{=}\sqrt{1{+}4x^2 M^2/Q^2}$, and $F_2^{(0)}(x,Q)$ is the structure
function in the limit where the proton mass $M$ is set to zero.
The additional terms, indicated by the dots, are given by
convolution integrals for which explicit expressions are given
in Eqs.~(22)-(24) of Ref.~\cite{Schienbein:2007gr}.
The magnitude of these corrections can be sizeable, especially for lower
scales, $Q\!\sim\! M$, 
as we illustrate in Fig.~\ref{fig:tmc} where we show
the $x$ dependence of the ratio $F_2^\mathrm{TMC}(x,Q)/F_2^\mathrm{{leading}\atop{TMC}}(x,Q)$ where
$F_2^\mathrm{{leading}\atop{TMC}}(x,Q)=x^2/(\xi^2 r^3)\, F_2^{(0)}(\xi,Q)$.
As can be seen, for $F_2$, the leading TMC gets modified by the additional TMC by up to a ${\sim}25\%$ at intermediate to 
large $x$ values. 

The leading TMC effects due to the Nachtmann variable $\xi$ and the pre-factor $x^2/(\xi^2 r^3)$
are  accounted for in all our calculations.
However, we have not included the effect of the convolution terms.
Since we are fitting only nuclear ratios such as $F_2^A/F_2^D$ and $\sigma_A/\sigma_D$, 
the nuclear $A$ dependence of these additive convolution terms is very minor.\footnote{%
The NMC $F_2^D$ data set ID=5160 is the one case which is not a ratio. 
This set extends out to $x{\lesssim}0.48$, and we have checked that these TMC 
corrections are less than the data uncertainty.
} %

One important question which arises is: do the TMCs scale as the mass of the
nucleus $M_A/Q$, or the mass of the nucleon $M/Q$ where $M_A= A M$?
This issue is especially important for the heavy nuclear targets such as
lead (A=208) where a $M_A/Q$ scaling would dramatically impact the low energy data.
The answer can be obtained by tracing the mass terms in the derivation of the master formula
(\textit{e.g.}, Eqs.~(22--24) of Ref.~\cite{Schienbein:2007gr}) from the OPE.
For a nucleon, the OPE expansion is in terms of $Mx/Q$ while for a full nucleus
this becomes $M_A x_A /Q$. Here, $x_A$ is the Bjorken $x$ for the nucleus
defined as $x_A=Q^2/(2 P_A\cdot Q)$, where $P_A$ is the full nucleus momentum
and  $x_A\in [0,1]$.
Thus, we find  $M_A x_A /Q = (A M)(x/A)/Q\equiv Mx/Q$, 
(where $x{=}A \, x_A$ and $x\in[0,A]$)
so even
for heavy nuclei the TMC terms are suppressed by powers of $M/Q$ where
$M{\sim}1$~GeV.
Consequently, the master formula of Ref.~\cite{Schienbein:2007gr} holds for both nucleons and nuclei.
There are certainly some subtle steps in extending the OPE and DGLAP formalism
from the case of a proton to a heavy nuclear target, and a separate paper
expanding the work of Ref.~\cite{Schienbein:2007gr} with
detailed derivations is in progress~\cite{tmc}.

Figure~\ref{fig:tmc} displays a band of curves for each fixed $Q$ value which show  the variation 
between~D (the lowest curve) and~Pb (the highest curve).
The spread of these bands can be as large as~5\% for $Q{=}1.0$~GeV, and decreases quickly for larger~$Q$ values. 
For the current investigation, we will focus on cuts of $Q{>}1.3$~GeV and $W{>}1.7$~GeV.
For example, from Fig.~\ref{fig:kin} we see that for $Q{=}1.3$~GeV,  the $W{>}1.7$~GeV cut includes points
out to $x\,{\lesssim}\,0.46$, and here the spread of this band (yellow) is on the order of ${\sim}1\%$.
This  represents the maximum deviation between a deuteron~(D) target and a lead~(Pb) target;
lighter nuclei such as carbon~(C) would be less. 
Therefore, we see that within our kinematic cuts, 
the TMC correction provides a uniform shift, within about~1\% or less, of all the nuclei.

Since   these additional TMC corrections shift both the $A$ and $D$ results by nearly the same factor,
the $F_2^A/F_2^D$ ratio will not be affected by the non-leading
TMCs:
\begin{equation}
    \frac{F_2^{A,\mathrm{TMC}}(x,Q)}{F_2^{D,\mathrm{TMC}}(x,Q)} \simeq
    \frac{F_2^{A,\mathrm{{leading}\atop{TMC}}}(x,Q)}{F_2^{D,\mathrm{{leading}\atop{TMC}}}(x,Q)}
    = \frac{F_2^{A,(0)}(\xi,Q)}{F_2^{D,(0)}(\xi,Q)}
    \, .
\end{equation}
Therefore, for our present study these leading-twist TMCs do not impact the results. 
However, if we were to examine the absolute structure functions (instead of ratios)
or the absolute cross-sections, these TMCs must be incorporated 
in the high $x$, low $Q$ region.%
\subsection{Higher Twist Corrections}
\label{sec:ht}

\begin{figure}[tb]
\includegraphics[width=0.49\textwidth]{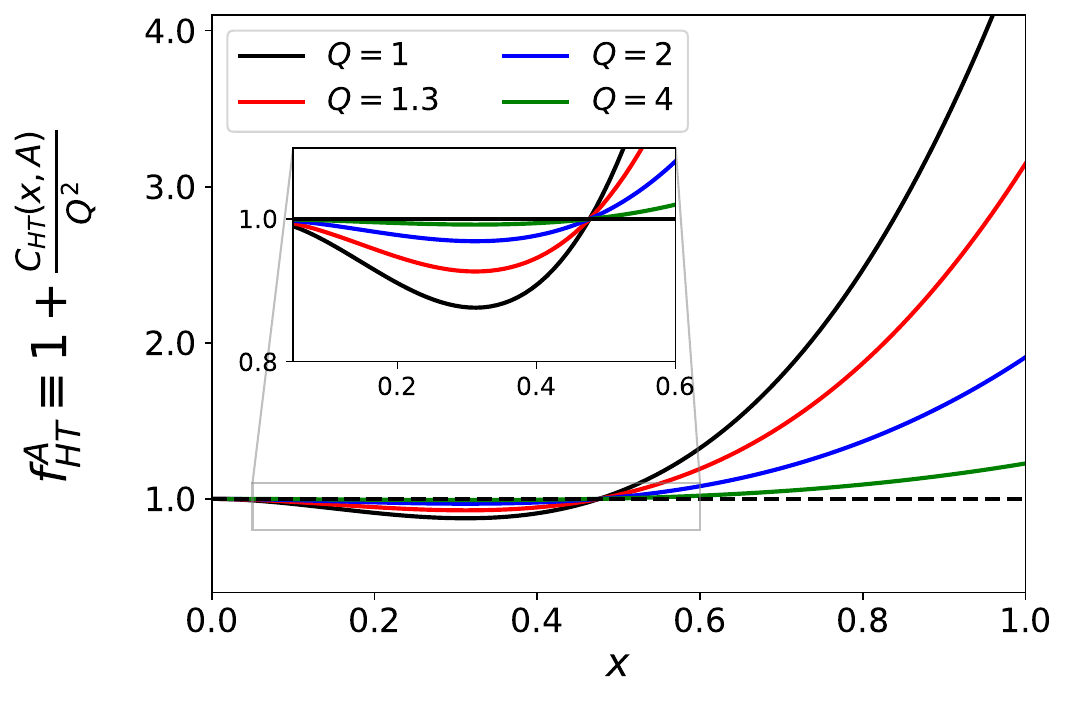}
\caption{We display the higher twist corrections as a function of $x$  for  $Q=\{1.0, 1.3,2,4\}$~GeV. 
The lower $Q$ values yield larger corrections since they scale as~$1/Q^2$.
}
\label{fig:ht}
\end{figure}

At high $x$ and low $Q$ values, the sub-leading $1/Q^2$ higher twist (HT) and  residual
power corrections can be important. 
To explore these effects, 
we will use a  phenomenological $x$-dependent function taken from the CJ15 study~\cite{Accardi:2016qay}
with the following form:
\begin{equation}
F_2^A(x,Q) \to  F_2^{A}(x,Q) 
\left[ 1+ \frac{C_{HT}(x,A)}{Q^2} \right]
    \quad , 
\label{eq:ht}
\end{equation}
where the higher-twist coefficient $C_{HT}$ depends on $x$ and the nuclear $A$:
\begin{equation}
C_{HT}(x,A) =  h_0 x^{h_1} (1+ h_2 x) \ A^{\tau} 
    \quad , 
\label{eq:ht2}
\end{equation}
with 
$\{h_0, h_1, h_2\}=$ \{-3.3~{\rm GeV}${}^2$, 1.9, -2.1\};
$h_0$ carries units of GeV${}^2$, and the $h_{1,2}$ are dimensionless.

Note that  the  $A^{\tau}$ scaling was not part of the original CJ15 formulation
and was included here to explore a possible nuclear dependence~\cite{Qiu:2001hj}. 
We have varied the $\tau$ exponent in our fits and the optimal values is indistinguishable 
from zero $(\tau{\sim}0\pm0.005)$; hence, we find no significant $A$ dependence for this correction. 
As we are primarily focused on lighter nuclei \{He, Be, C\} in this study,
it may be useful  to revisit the $A^{\tau}$ dependence for the case of   heavy nuclei.

Also note that since the $h_0$ term in Eq.~(\ref{eq:ht2}) will generally be scaled by the target mass $M$,
the separation between the higher-twist and the Target Mass Corrections (TMCs) discussed above is not unambiguous.

In Fig.~\ref{fig:ht} we display the higher twist  correction as 
a function of $x$   for a selection of $Q$ values. 
The HT correction gives a slight reduction in the intermediate $x$ region ($x{\sim} 0.3$), 
and then becomes large and positive at large $x$, depending on the $Q$ value. 
For example, at \mbox{$x=0.7$}, the correction factor is ${\sim}1.5$ for the \mbox{$Q=1.3$~GeV,} 
whereas at  \mbox{$Q=2$~GeV} this is already reduced to   ${\sim} 1.2$.
In Fig.~\ref{fig:ht} we also provide an inset plot to show the details in the intermediate $x$ region.\footnote{%
Note for the neutrino community.  Early theoretical considerations~\cite{Shuryak:1981kj,Luttrell:1981ke}  of higher-twist (HT) effects indicated there could be real differences in these effects between weak and electromagnetic scattering.  Indeed, early experimental investigations~\cite{Morfin:1981rn,Bosetti:1978kz,Varvell:1987qu} of higher twist in neutrino-nucleon scattering support this conclusion and indicate that the HT contribution does not become positive at higher-$x$ as in Fig.~\ref{fig:ht}, but remains negative in the high $x$, lower $Q^2$ region.  This neutrino HT behavior needs further exploration, however it suggests consideration of this difference in applying  the results of this study to neutrino scattering in the relevant hi-$x$, low $Q^2$ region.
} %

\goodbreak
\subsection{Deuteron Nuclear Structure}\label{sec:deut}
\nobreak

A large fraction of the nuclear DIS data incorporated in global nPDF fits, including the present one, are  expressed as a $F_2^A/F_2^D$ ratio of structure functions on a heavy nuclear target and on a deuteron target, see Tables~\ref{tab:jlab}-\ref{tab:exp3} in the Appendix. A careful treatment of the denominator is therefore called for in order to analyze the effects of the nuclear dynamics on the nuclear quark and gluons, as most recently highlighted in Ref.~\cite{Szumila-Vance:2020zpt}. 
 
The deuteron is the lightest and least bound of all compound nuclei. Therefore it is typically considered as composed of a free proton and a free neutron, and the deuteron structure function is computed as the sum of the structure functions of its components. Extensive studies, however, indicate that the deuteron structure is far more complex than a simple combination of proton and neutron PDFs
might suggest \cite{Owens:2007kp,Owens:2012bv,Malace:2014uea,Accardi:2016qay,Alekhin:2017fpf,Ball:2020xqw,Accardi:2021ysh}. 
Yet, the deuteron differs from the heavier targets considered in this paper because its binding energy is far smaller, making this a loosely bound system of two nucleons rather than a tightly bound and noticeably denser system of many nucleons interacting with each other \cite{Accardi:2016muk}. Thus its nPDFs may not follow the $A$-scaling parametrization utilized in this work. 

For these reasons\,---\,and to avoid double counting with the proton PDF fits that we use to anchor the nuclear PDFs parametrization used in this paper which already include deuteron target data\,---\,we choose not to fit the deuteron nPDF. Instead, we build upon the comprehensive studies available in the literature, and calculate the deuteron structure function, using the PDFs and nuclear dynamics simultaneously extracted in the recent CJ15 global QCD analysis~\cite{Accardi:2016qay}.
At large $x$, this nuclear dynamics calculation is based on  two main components: 
i)~a baryonic smearing function evaluated on the basis of the AV18 nuclear wave function \cite{Wiringa:1994wb,Veerasamy:2011ak} that accounts for Fermi motion and nuclear binding effects,
and
ii)~a fitted parametric function quantifying the effect of the offshellness of the bound nucleon. At smaller $x$, shadowing effects are calculated according to Ref.~\cite{Melnitchouk:1992eu}.

\begin{figure}[tb]
\begin{subfigure}
\centering
 \begin{subfigure}{a)}
\includegraphics[width=0.435\textwidth]{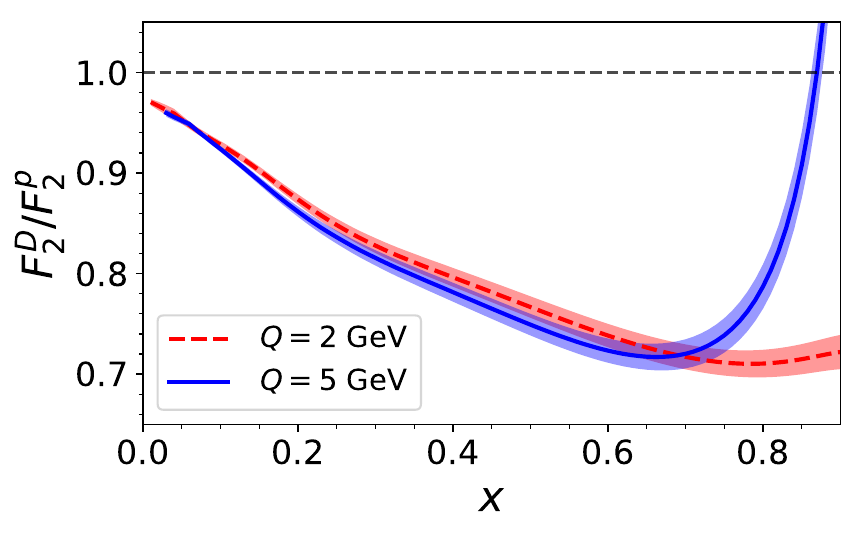}
  \end{subfigure}
\hfil
 \begin{subfigure}{b)}
\includegraphics[width=0.45\textwidth]{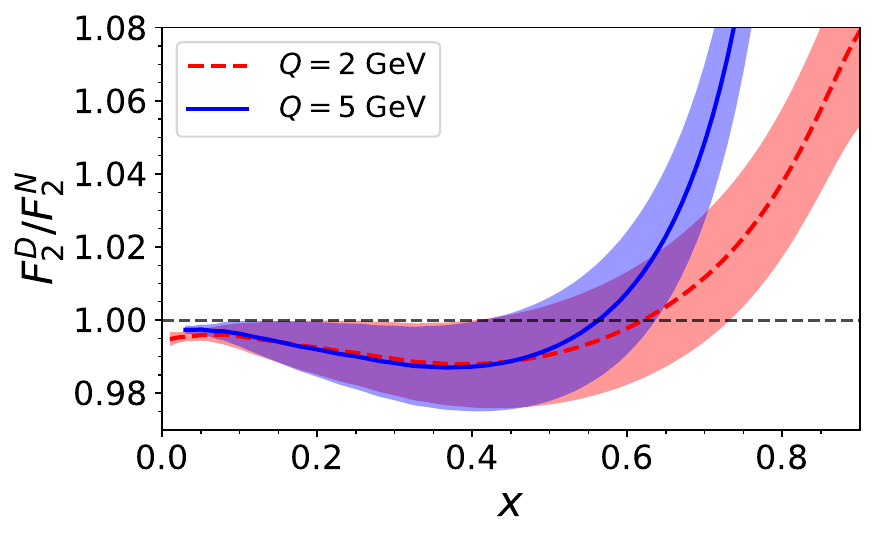}
  \end{subfigure}
\caption{We display the CJ15 calculation (Ref.~\cite{Accardi:2016qay}) of the structure function ratio of the deuteron $F_2^D$ to the free proton $F_2^p$
as a function of $x$  for selected values of $Q$. 
Note that these curves are effectively fits to the experimental data.
Additionally we display the ratio of $F_2^D$ to the isoscalar combination $F_2^N = F_2^p+F_2^n$
to demonstrate the nuclear dynamics in the deuteron beyond isospin symmetry effects. 
The determination of the neutron structure function in the denominator depends on the details of the nuclear correction model fitted in Ref.~\cite{Accardi:2016qay}.
The colored bands are 90\% confidence level uncertainty arising from the global CJ15 fit parameters.
} %
\label{fig:deut}
\end{subfigure}
\end{figure}

In Fig.~\ref{fig:deut} we display 
the ratio of $F_2^D$ to the proton's $F_2^p$, which we use in the present nPDF analysis.
For comparison, we also display   the ratio
to the isoscalar combination, $F_2^N=F_2^p+F_2^n$, which illustrates 
the size and $x$ dependence of the nuclear effects in the deuteron system, after removing the underlying isospin symmetry effects.
We can see that the deuteron-to-proton ratio, $F_2^D/F_2^p$, dips to ${\sim}70\%$
at $x{\sim}0.7$ before dramatically increasing at larger $x$.
Comparing this with the ratio to the isoscalar structure function, $F_2^D/F_2^N$,
we see much of the previous effect was in fact due to the differing charge
factors weighting the quark PDFs in the isoscalar combination, but the residual dynamical
nuclear effects are not negligible, particularly at large-$x$
values.

Note that the correct treatment of the deuteron structure function will impact, in principle, a large fraction of the DIS data, see Tables~\ref{tab:jlab}-\ref{tab:exp2}.
The nuclear dynamics in the deuteron may reasonably be neglected in typical nPDF fits because, after the usual kinematic cut on $W\gtrsim 3.5$ GeV, most of the DIS data lies in a small-to-medium $x$ region where the corrections are small.
However, if we want to extend our fit into the large $x$
region, Fig.~\ref{fig:deut} demonstrates that it is essential %
to go beyond pure isospin symmetry considerations.

Operationally, we incorporate the deuteron's nuclear dynamics by rescaling the $F_2^A/F_2^D$ ratio data by a $F_2^D/F_2^p$  factor calculated using the CJ15 PDFs and deuteron correction model (Fig.~\ref{fig:deut}a).
Thus, the transformation is:
\begin{align}
\frac{F_2^A}{F_2^p} \equiv \frac{F_2^A}{F_2^D} \cdot \left( \frac{F_2^D}{F_2^p} \right)_{CJ}
\quad , 
\label{eq:dCorr}
\end{align}
and this effectively produces a 
nucleus-to-proton ratio data set that is then passed to the fit 
program.
We provide the data and associated corrections in our supplementary material.

Note that the $\left( F_2^D/F_2^p \right)_{CJ}$ rescaling factor is computed 
using the structure function results of the CJ15 global analysis~\cite{Accardi:2016qay,Accardi:2016muk}.
CJ15 includes data from a variety of processes including 
DIS, Drell-Yan and jets from hadron-hadron collisions at the Tevatron; this combination 
covers a broad kinematic range which contributes to improved precision. 
Additionally CJ15 includes data from the BONuS experiment~\cite{Tkachenko:2014byy,Baillie:2011za}
which measured the  neutron to deuteron 
$F_2$  structure function ratio using an innovative spectator tagging technique.
The BONuS data provides unique information on the neutron structure function,
and this  allows substantive improvement of the up and down quark flavor separation
in the CJ15 analysis.

In principle we could also use CJ15 to correct to an isoscalar quantity (\textit{e.g.}, $F_2^D/F_2^N$)
as illustrated in \mbox{Fig.~\ref{fig:deut}-b).} 
The difference between this approach and Eq.~\eqref{eq:dCorr} depend 
on the details of the CJ15 PDFs, including the specific $F_2$ neutron structure function extracted.
For this reason, we have chosen to implement Eq.~\eqref{eq:dCorr}; but, 
we have verified that this choice does not substantively impact 
our conclusions for the valence distributions, which is the focus of this study.

As in the DIS case, the $p{+}D$ Drell-Yan cross section is modified by the nuclear dynamics in the deuteron target compared to calculations that treat the deuteron as composed of a free proton and a free neutron. The nuclear smearing model discussed above can also be applied to this case \cite{Ehlers:2014jpa}. However, we have verified that the ensuing deuteron corrections amount to less than 1\% in the kinematics covered by the available DY data, and can be neglected compared to the experimental uncertainties.

\section{Fits to JLab DIS Data Sets}
\label{sec:analysis}

Now that we have surveyed the various phenomena which enter the fit
in the large $x$ and low $Q$ region, we will implement these effects into 
our fit so we can compare the relative effects.

In this section, we will first review the experimental data that enters the fit,
and then provide an overview of the exploratory fits we will use to 
investigate the effects discussed in the previous section.

\subsection{The Experimental Data }
\label{sec:data}
Historically many experiments have included isoscalar corrections (accounting
a for different number of protons and neutrons in nuclei) in their data,
e.g. structure function ratio data~\cite{Gomez:1993ri,Ashman:1992kv}. Since these corrections
are model dependent, for the purpose of the current analysis, we have removed all
isoscalar corrections from the used data sets following the specific prescriptions
used by the collaborations.
We provide supplemental materials with the tabulated data used in our analysis 
along with these corrections. This is available in a separate text file. 

\subsubsection{Data Used in nCTEQ15}
\label{sec:ncteqdata}

For our fits, we will use the full set of DIS and DY data from the nCTEQ15 analysis~\cite{Kovarik:2015cma} 
with isoscalar corrections removed. 
These data are listed in Tables~\ref{tab:exp1}-\ref{tab:exp3} in Appendix~\ref{sec:datatabs}.

Additionally, the nCTEQ15 analysis includes RHIC inclusive pion data from PHENIX~\cite{Adler:2006wg}
and STAR~\cite{Abelev:2009hx} collaborations for nuclear modification $R_{\mathrm{DAu}}^{\pi}$.
The pion data helps to constrain the nuclear gluon PDF, especially in the lower $x$ regions. 
As our focus will be on the high-$x$ region, we have not included the pion data in the current fits
which simplifies our analysis as we don't need to consider fragmentation functions. 
However, as a cross check, we separately compute the $\chi^2$ values for the pion data to ensure
our results are compatible with these data sets. These comparisons are 
reported in Sec.~\ref{sec:uncert} and show very good agreement with these data.

\subsubsection{The JLab Data}
\label{sec:jlabdata}

In this analysis, we include new DIS data from the Jefferson Lab
CLAS~\cite{Schmookler:2019nvf}  and Hall~C~\cite{Seely:2009gt} 
experiments taken during the 6~GeV electron beam operation \cite{JLab6-review:2011}.
These data sets, in addition to spanning a wide range of $A$, 
provide high-precision constraints for nuclear PDFs at  high-$x$ and low-$Q^2$. 
Nuclei included are ${}^3$He, ${}^4$He, ${}^9$Be, ${}^{12}$C, 
${}^{27}$Al, ${}^{56}$Fe, ${}^{64}$Cu, ${}^{197}$Au, ${}^{208}$Pb, spanning $x$ from $0.2$ to $0.95$. 
Full details of these data sets are summarized in Appendix~\ref{sec:datatabs} 
as well as in Refs.~\cite{Schmookler:2019nvf,Seely:2009gt}.

\subsection{The Parameters}
\label{sec:parms}

As before, our nCTEQ PDFs are parameterized  as~\cite{Kovarik:2015cma}:
\begin{equation}
    x f_i^{p/A}(x,Q_0) = 
    c_0 x^{c_1} (1-x)^{c_2} e^{c_3 x} (1+ e^{c_4} x)^{c_5}
    \quad ,
\end{equation}
where the nuclear $A$ dependence is encoded in the $c_k$ coefficients 
as:\footnote{%
Note that the original nCTEQ15 parameterization of Ref.~\cite{Kovarik:2015cma}
used the notation $\{ c_{k,0}, c_{k,1}, c_{k,2} \}$
in place of the current $\{ p_k, a_k, b_k\}$ notation of Eq.~(\ref{eq:ck})
} 
\begin{equation}
    c_k \longrightarrow c_k(A) \equiv p_{k} + a_k (1-A^{-b_k})
    \quad ,
    \label{eq:ck}
\end{equation}
where $k{=}\{1, ... ,5\}$.
This parameterization has the advantage that in the limit $A{\to} 1$, $c_k(A)$ reduces to proton value $p_k$.
The $a_k$ parameters control the magnitude of the modification to the $p_k$ proton ``boundary condition,''
and the $b_k$ parameters control the power of the nuclear-$A$ term.
Specifically, within the nCTEQ15 framework we parameterize the combinations 
$\{u_v,d_v,(\bar{u}{+}\bar{d}),(\bar{d}/\bar{u}),s,g\}$,
and we impose the boundary condition\footnote{%
Note, that the recent \ncteqwz{}~\cite{Kusina:2020lyz} PDF set allows  additional freedom 
for the strange PDF as compared to the original nCTEQ15 set~\cite{Kovarik:2015cma}.
} %
of $s{=}\bar{s}{=}{\kappa (\bar{u}{+}\bar{d})/2}$
with $\kappa{=}0.5$ at $A{=}1$.

We will then perform fits using a total of 19 free parameters.
This includes 5 parameters for each the up and down valence
$\{ 
a_k^{u_v},
a_k^{d_v}
\}$ where $k{=}\{1 ... 5\}$, 
two parameters for the $\bar{u}+\bar{d}$ combination 
$\{ a_1^{\bar{d}+\bar{u}}, a_2^{\bar{d}+\bar{u}} \}$,
and seven parameters for the gluon 
$\{ 
a_{1}^{g}, 
a_{4}^{g}, 
a_{5}^{g}, 
b_{0}^{g}, 
b_{1}^{g}, 
b_{4}^{g}, 
b_{5}^{g}
\}$.
This set of 19 parameters includes  the original  16 parameters used in  the nCTEQ15 PDF fit,
and adds the  3 parameters $\{ a_3^{u_v}, a_3^{d_v}, a_4^{d_v}\}$
so that we have all the $a_k$ parameters open for the up and down valence 
to fully explore the parameter space. 
While our focus is on the high-$x$ region where the up- and
down-quark PDFs dominate, there is some interplay between the 
sea quarks and the gluon PDFs; hence, this will influence the $g$ and $\bar{d}+\bar{u}$ parameters.

For our analysis we use our new  \ncteqpp{} framework which has a modular structure
and links to a variety of external tools including
a modified version of HOPPET~\cite{Salam:2008qg}
(extended to accommodate grids of multiple nuclei),
APPLgrid~\cite{Carli:2010rw}, and MCFM~\cite{Campbell:2015qma}.
Additionally, we have used FEWZ~\cite{Li:2012wna} for benchmarking our WZ calculations,
and xFitter~\cite{Alekhin:2014irh} for various cross checks.

\newcommand{\baseChi}{1525}
\newcommand{\baseChiDOF}{0.99}
\newcommand{\htChi}{1489}  %
\newcommand{\htChiDOF}{0.96}  %
\newcommand{\deuChi}{1331}   %
\newcommand{\deuChiDOF}{0.86}  %
\newcommand{\ncteqHixChi}{1297}  %
\newcommand{\ncteqHixChiDOF}{0.84}  %
\begin{table}[tb]
\renewcommand{\arraystretch}{1.0}
\setlength{\tabcolsep}{6pt}
\begin{tabular}{|l|c|c|c|c|c|}
 \hline 
Fit & $\chi^2$ & $N_{data}$ &  $\nicefrac{\chi^2}{N_{dof}}$  & $Q_{cut}$ &$W_{cut}$\\ \hline \hline
nCTEQ15 & 587 & 740 & 0.81 & 2.0 & 3.5 \\ \hline 
nCTEQ15* & 2664 & 1564 & 1.70 & 1.3 & 1.7 \\ \hline  \hline
BASE & \baseChi{}  & 1564 & \baseChiDOF{} & 1.3 & 1.7 \\ \hline 
HT  & \htChi{} & 1564 & \htChiDOF{} & 1.3 & 1.7 \\ \hline 
DEUT & \deuChi{} & 1564 & \deuChiDOF{} & 1.3 & 1.7 \\ \hline 
\ncteqhix{} & \ncteqHixChi{} & 1564 & \ncteqHixChiDOF{} & 1.3 & 1.7    \\ \hline 
\end{tabular}
\caption{We summarize the fits explored in this study. 
Note the nCTEQ15 and nCTEQ15* results are not  re-fits; 
the first entry, nCTEQ15, lists the fit results from the original analysis, 
and the second entry, nCTEQ15*,  simply computes the $\chi^2$ for the new JLab data with 
the relaxed $\{Q, W \}$ cuts. 
The remaining four entries, \{BASE, HT, DEUT, \ncteqhix{}\}, 
are re-fit with the new data, and are detailed in the text.
}
\label{tab:chi2}
\end{table}
\begin{figure*}[t]
\includegraphics[width=0.95\textwidth]{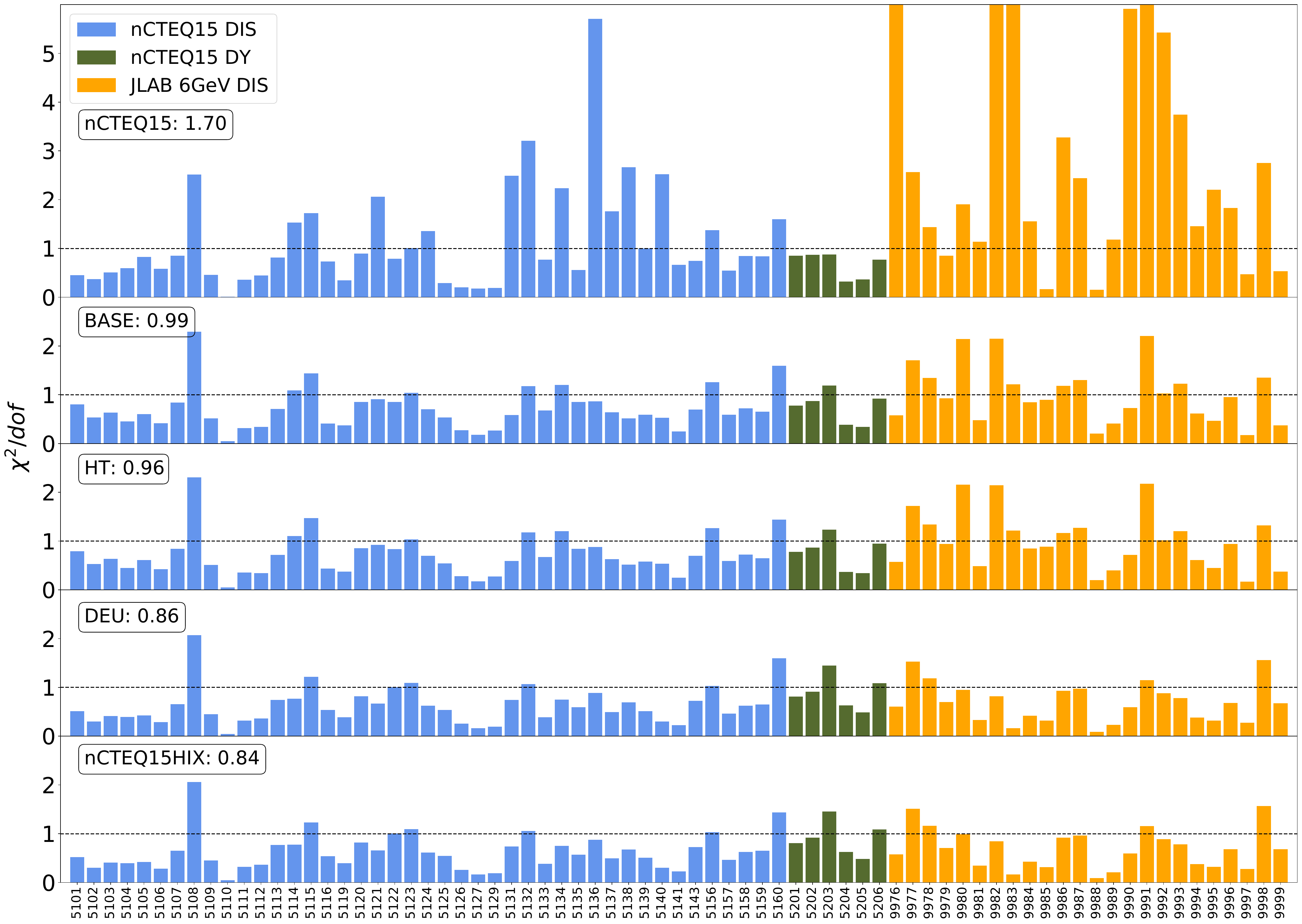}
\caption{The $\chidof$ for each individual experiment. 
The experiment ID is listed on the abscissa and are identified 
in Tables~\ref{tab:jlab}, \ref{tab:exp1}, \ref{tab:exp2} and~\ref{tab:exp3}.
The original DIS data used in the nCTEQ15 fit are in blue, the DY data in green, 
and the new JLab data in yellow.
We also list the total $\chidof$ for each fit inset in the plot. 
}
\label{fig:chidof}
\end{figure*}
\subsection{The Fits}
\label{sec:fits}

Having specified the required inputs, we now present a brief overview of the fits 
we will investigate. 
These are  summarized in Table~\ref{tab:chi2} and are outlined in the text below.

\begin{description} %

\item[{\bf nCTEQ15*}] This is the original set of nuclear PDFs
  as computed in Ref.~\cite{Kovarik:2015cma}. The new data has \textbf{not} been fit; we have simply computed the  $\chi^2$ values.

\item[{\bf BASE}] This BASE fit will serve as our primary reference fit
as it simply extends the nCTEQ15 fit by reducing the kinematic cuts to 
$Q{>}1.3$~GeV and $W{>}1.7$~GeV. 
In particular, this fit does not include any %
higher-twist corrections, or deuteron modifications.

\item[{\bf HT}] 
    This fit extends the BASE fit by adding the higher twist (HT) power corrections   as described in Eq.~(\ref{eq:ht}).
    
\item[{\bf DEUT}] 
    This fit extends the BASE fit by including the deuteron modifications  as described in Sec.~\ref{sec:deut}.
    
\item[{\bf \ncteqhix{}}] 
    This fit extends the BASE fit by including both 
    the deuteron correction  (DEUT) and also  
    the higher twist (HT) corrections.

\end{description}
\noindent
Note, that all of the above fits include the TMCs originating
from the scaling variable $\xi$ and from the prefactor, as explained
in the TMC discussion, Sec.~\ref{sec:tmc}.

Table~\ref{tab:chi2} also  lists the $\{Q,W\}$ cuts used in each fit. 
The original nCTEQ15 fit used cuts of $Q{>}2.0~{\rm GeV}$, and  $W{>}3.5~{\rm GeV}$.
For the new fits, we will relax these cuts  to 
$Q{>}1.3~{\rm GeV}$, and  $W{>}1.7~{\rm GeV}$ to expand the kinematic range 
covered.

The $W{>}1.7~{\rm GeV}$ cut is chosen to avoid 
resonance contributions.
However, there are indications that it may be possible to further reduce this cut
to extend coverage of the  Shallow Inelastic Scattering (SIS) region~\cite{SajjadAthar:2020nvy}, and
we examine this briefly in Sec.~\ref{sec:noWcut}.

Now that we have outlined the fits,
we present the results in the following section. 

\section{Results of the Fits}
\label{sec:results}
\subsection{The \texorpdfstring{$\chi^2$}{chi2}  and \texorpdfstring{$\chidof$}{chi2/dof} Values}
\label{sec:chi2}

The  $\chidof$ values provide a succinct measure of the quality of the fit,
and these are summarized in Table~\ref{tab:chi2}. 
Additionally,  Fig.~\ref{fig:chidof} displays the $\chidof$ 
for each individual experiment for the separate fits, and the 
experiment IDs are listed in Tables~\ref{tab:jlab}, \ref{tab:exp1}, \ref{tab:exp2}, and~\ref{tab:exp3}.

{\bf nCTEQ15*:} Starting with {nCTEQ15}, we first present the values for the 
original fit which used cuts of
$Q{>}2$~GeV and $W{>}3.5$~GeV  and obtained 
an excellent $\chidof$ of 0.81.
However, if we take this nPDF and compute the $\chidof$ 
including the new data (and new less stringent cuts)---without fitting---we obtain a very large $\chidof$  of~1.70. 
Inspecting Fig.~\ref{fig:chidof} we see that this large value
reflects the fact that these new data, which are beyond the 
kinematic bounds of the original fit, are not well described.%
    \footnote{The effect of removing isoscalar corrections in the
    original nCTEQ15 data set is limited.}
These two entries in Table~\ref{tab:chi2}  provide a useful benchmark so we can see how our new fits improve in comparison. %

{\bf BASE:} 
The {BASE} fit will serve as our reference. 
In this fit, we have simply taken nCTEQ15 as a starting point,
and preformed a new fit including the JLab data; no additional corrections are included. 
Comparing to nCTEQ15*, we see  substantial improvement as the 
$\chi^2$ reduces from 2664 to \baseChi{} to yield a $\chidof$ of \baseChiDOF{}. 
In Fig.~\ref{fig:chidof} we observe that the $\chidof$ 
values for the individual experiments have also improved significantly.

{\bf HT:} 
The {HT} fit applies the higher-twist correction of Sec.~\ref{sec:ht},
and  yields moderate improvement compared to the BASE fit (\htChi{} vs.\ \baseChi{}), 
as well as associated improvements
in the individual experiments of Fig.~\ref{fig:chidof}.

{\bf DEUT:} 
The {DEUT} fit applies the deuteron correction of Sec.~\ref{sec:deut},
and  yields significant improvement compared to BASE (\deuChi{} vs.\ \baseChi{}).
Additionally, in Fig.~\ref{fig:chidof} we observe that 
the fit to the JLab data sets has improved and all the individual $\chidof$ values 
are now below 2.0, in contrast to both the BASE and HT fits.

{\bf \ncteqhix:}
Finally, the {\ncteqhix{}} fit applies both the deuteron correction 
and the higher twist power correction. This combination 
yields the best fit (\ncteqHixChi{} vs.\ \baseChi{}), and reduces the $\chidof$  to $\ncteqHixChiDOF{}$. 
Furthermore, examining the   $\chidof$ values  
of Fig.~\ref{fig:chidof} we see each experiment is well fit,
with the exception of set 5108 which is a known outlier.\footnote{%
  Note, we find the DIS experiment 5108 (Sn/D EMC-1988) 
  to be an outlier with  $\chidof>2$, and this is consistent with other
  analyses~\cite{Eskola:2016oht,deFlorian:2011fp}.
}
Comparing to the BASE fit, 
we  observe that some of the original nCTEQ15 DIS data sets show a reduced $\chidof$ 
indicating the HT and DEUT corrections also improve the fit to these data.

{\bf Summary:} 
In summary, we find that the HT modifications provide some improvements  
(${\sim}3\%$ of $\chidof$),
the DEUT modifications provide larger improvements  
(${\sim}10\%$ of $\chidof$),
and the combination of the HT and DEUT modifications yield the 
best fit with an improvement of  ${\sim}15\%$ of $\chidof$.

\begin{figure*}[t]
\centering
\begin{subfigure}
\centering
\includegraphics[width=0.92\textwidth]{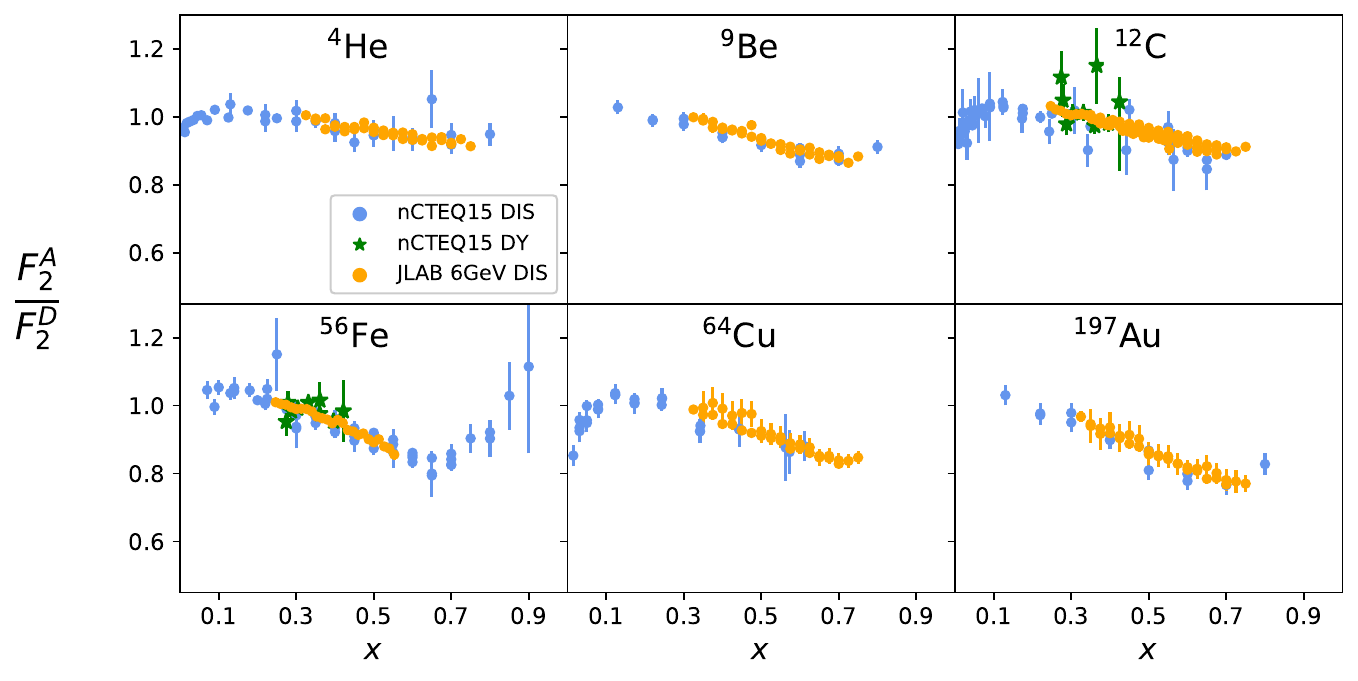}
\caption{We display the $F_2^A/F_2^D$ ratio  of selected data sets sorted by nuclei. 
The data from the original nCTEQ15 DIS points are in blue, and the DY in green.
The new JLab DIS data are in yellow. 
 }
\label{fig:sample1a}
\end{subfigure}
\begin{subfigure}
\centering
\includegraphics[width=0.92\textwidth]{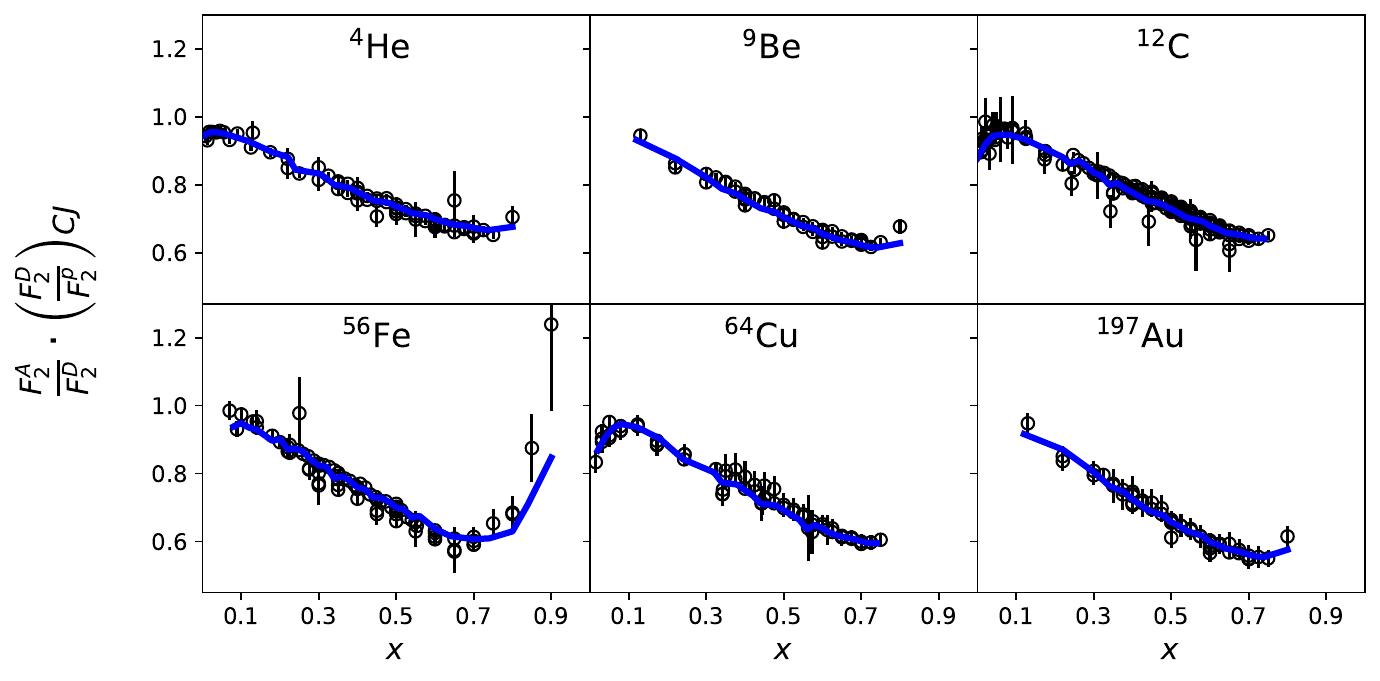}
\caption{We display the $F_2^A/F_2^D \cdot \left( F_2^D/F_2^p \right)_{CJ} $ ratio 
for selected data sets sorted by nuclei. 
We also overlay the theoretical prediction of \ncteqhix{} in blue. 
The theory predictions have
been calculated at averaged $Q$ values where data sets overlap.
}
\label{fig:sample1b}
\end{subfigure}
\end{figure*}
\subsection{Data and Theory Comparison }
\label{sec:datatheory}

Having evaluated the $\chidof$ for the separate experiments, 
we will examine some sample comparisons of our predictions with the data. 

In Figure~\ref{fig:sample1a} we display the nucleus-to-deuteron $F_2^A/F_2^D$ ratio for the 
selected  data sets sorted by nuclei. The characteristic EMC shape is evident, as well as the 
A dependence.
For example, we see that starting at $x{\sim}0.3$ where the ratio is ${\sim}1.0$ for all A, by $x{\sim}0.7$ this ratio dips to 
${\sim}0.95$ for  ${}^4$He,
and to  ${\sim}0.75$ for  ${}^{197}$Au.

In Figure~\ref{fig:sample1b}  we display the nucleus-to-proton $(F_2^A/F_2^D) \cdot (F_2^D/F_2^p)_{CJ}$ 
ratio, again sorted by nuclei. Here, we have multiplied by the ratio $(F_2^D/F_2^p)_{CJ}$ 
taken from the CJ15 study~\cite{Accardi:2016qay}, shown in Fig.~\ref{fig:deut}, to approximately convert
the results of the previous figure to ${\sim}(F_2^A/F_2^p)$. 
Note that the introduction of the $x$-dependent multiplicative $(F_2^D/F_2^p)_{CJ}$ factor
visually suppresses the $A$-dependent change in slope seen in Figure~\ref{fig:sample1a}. However, a check
of the values of $(F_2^A/F_2^p)$ at $x{\sim}0.3$ and $x{\sim}0.7$ for ${}^4$He and ${}^{197}$Au confirms
that the $A$-dependent change in slope has been maintained.

In Fig.~\ref{fig:sample1b}   we also display the corresponding
theoretical 
calculations (blue line) obtained with the \ncteqhix{} PDFs. We can see 
that they provide a very good description of the fitted data.

\begin{figure*}[thp]
\begin{subfigure}{}
\null\vspace{-0.15in}
\begin{center}
{
\includegraphics[width=0.95\textwidth]{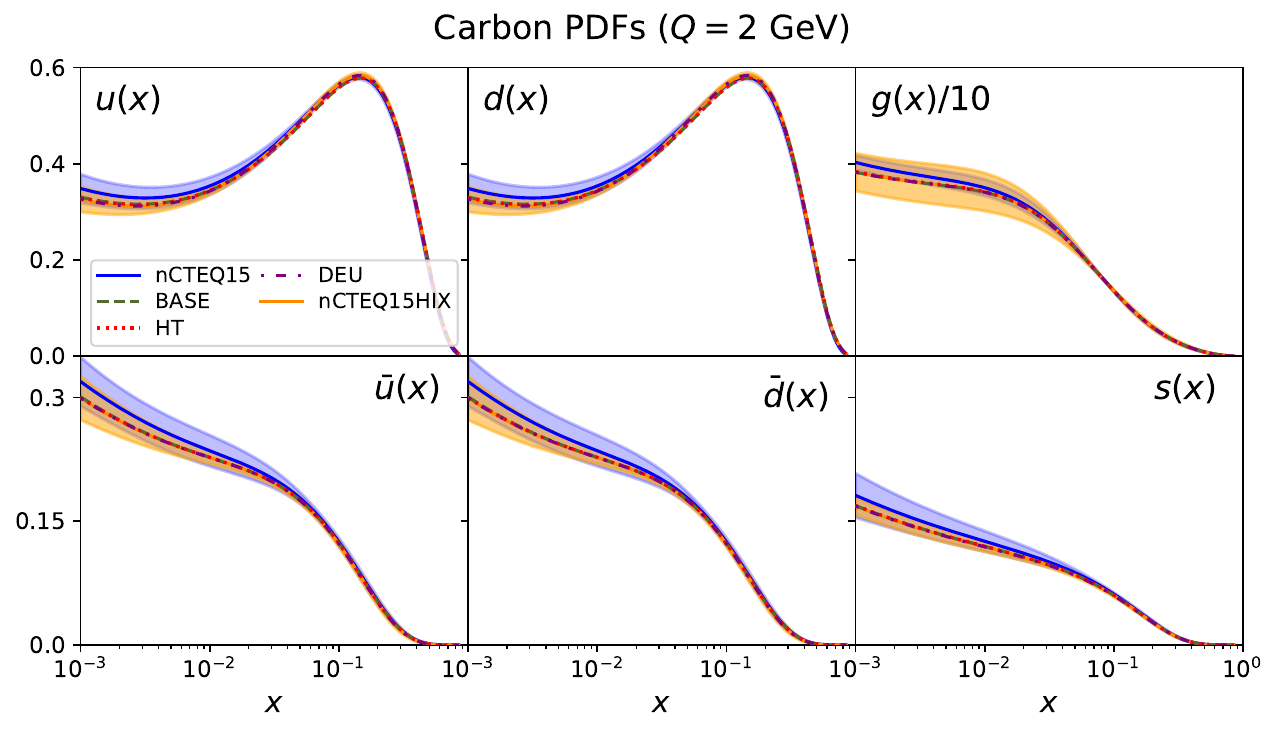}
}
\end{center}
\null\vspace{-0.20in}
\caption{Carbon (${}^{12}$C) nPDFs $x f^{\mathrm{C}}(x,Q)$ at $Q=2$~GeV.
We show the uncertainty bands for nCTEQ15 (blue) and \ncteqhix{} (yellow) 
computed with the Hessian method. 
} %
\label{fig:carbon1}
\end{subfigure}
\begin{subfigure}{}
\begin{center}
{
\includegraphics[width=0.95\textwidth]{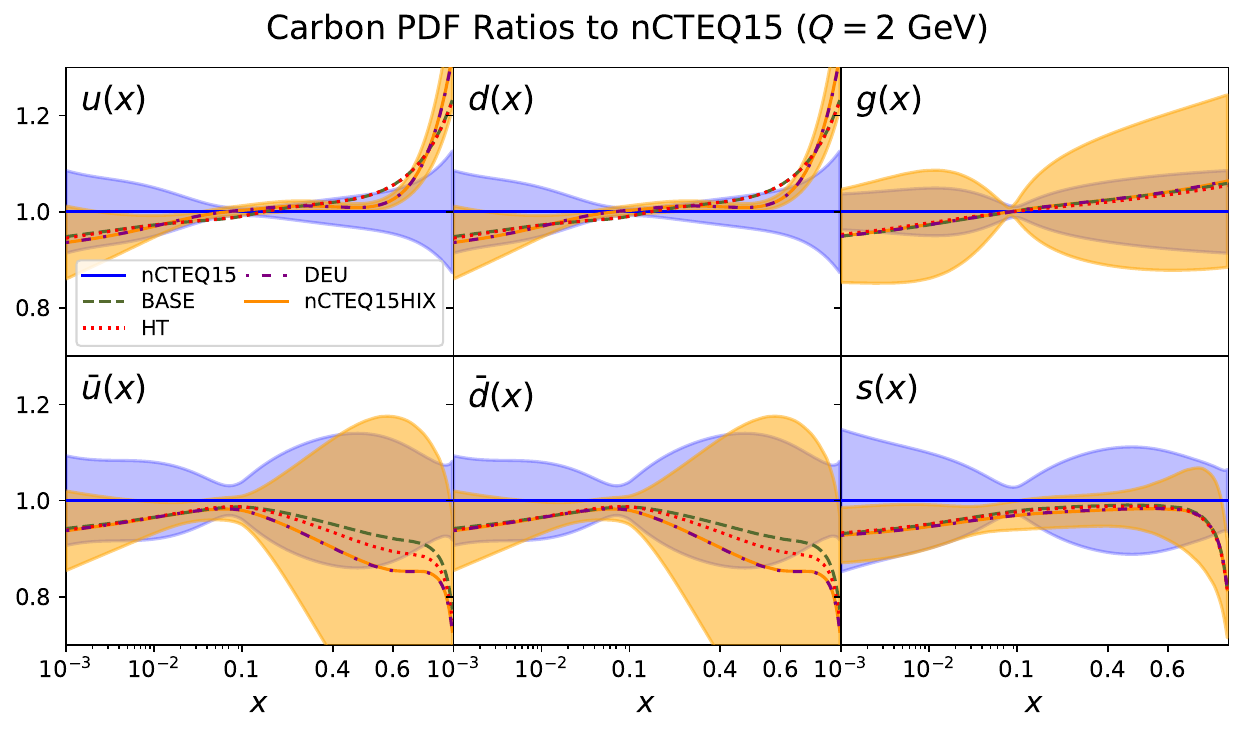}
}
\null\vspace{-0.10in}
\caption{%
The corresponding ratio of nPDFs compared to the nCTEQ15 central nPDFs for ${}^{12}$C
using a log-linear scale to highlight the large-$x$ region.
We show the uncertainty bands for nCTEQ15 (blue) and \ncteqhix{} (yellow) 
computed with the Hessian method. 
Note that while DEUT and \ncteqhix{} are distinct nPDFs which yield differing $\chidof$ values,
these differences are imperceptible  on the scale of this figure, as well as in  Figs.~\ref{fig:iron1}--\ref{fig:lead1}.
}
\label{fig:carbon2}
\end{center}
\end{subfigure}
\end{figure*}
\begin{figure*}[tb]
\includegraphics[width=0.95\textwidth]{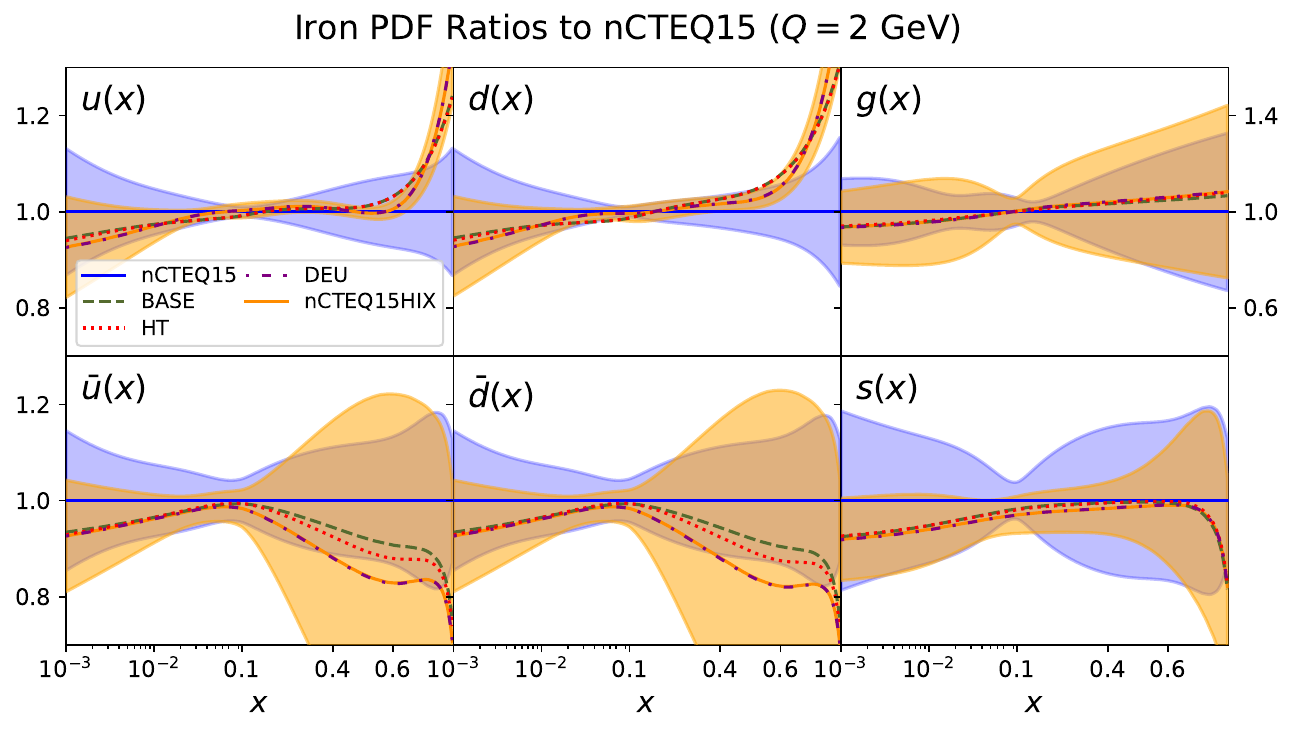}
\caption{Iron (${}^{56}$Fe) PDFs ratio compared to nCTEQ15 at $Q=2$~GeV.
We show the uncertainty bands for nCTEQ15 (blue) and \ncteqhix{} (yellow) 
computed with the Hessian method. 
}
\label{fig:iron1}
\end{figure*}
\begin{figure*}[tb]
\includegraphics[width=0.95\textwidth]{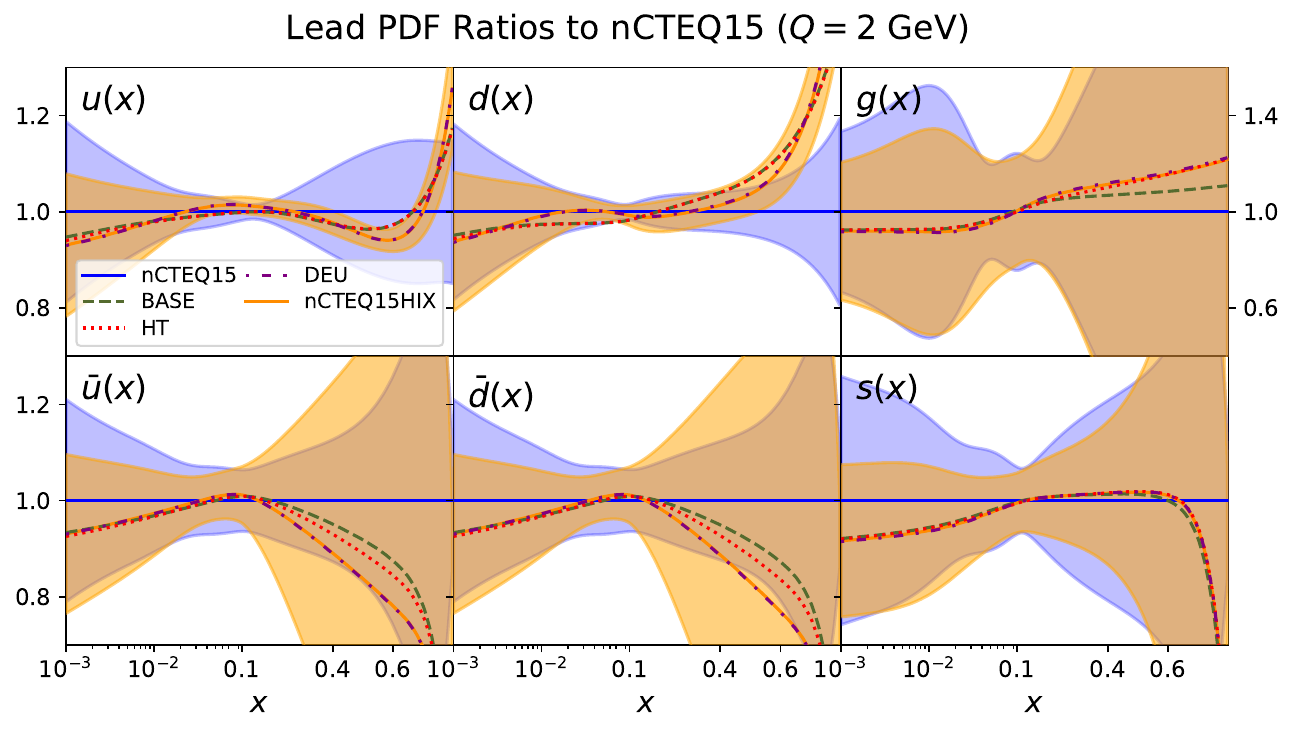}
\caption{Lead (${}^{208}$Pb)  PDFs ratios compared to nCTEQ15 at $Q=2$~GeV. 
We show the uncertainty bands for nCTEQ15 (blue) and \ncteqhix{} (yellow) 
computed with the Hessian method. 
}
\label{fig:lead1}
\end{figure*}
\subsection{The PDFs} \label{sec:pdfs}

Finally, we examine the impact of the various corrections on 
the underlying PDFs. 
Although the up and down flavors dominate in the high-$x$ region, 
these PDFs will also influence the gluon and sea-quark distributions indirectly. 
Thus, we must examine all the flavors to obtain a complete picture of the impact of 
the JLab data. 

We will present a sample of three representative nuclei: 
\{${}^{12}$C,   ${}^{56}$Fe,  ${}^{208}$Pb\}.
As the bulk of the new data is at lower $A$ values we will focus our attention on the 
${}^{12}$C PDFs, but we will want to examine the heavier nuclei to infer the
various trends in $A$ as well as to verify that our extrapolations to Pb are reasonable. 
For these plots, we have also displayed the uncertainty bands computed
with the Hessian method for both  \ncteqhix{} and the original nCTEQ15 PDF sets. 
Note, that when plotting nuclear PDFs, we always display the full
nuclear distribution, $f^A$, constructed out of the bound proton (neutron)
distributions, $f^{p(n)/A}$, as: 
\begin{equation}
f^A = \frac{Z}{A}\ f^{p/A} + \frac{(A-Z)}{A}\ f^{n/A} \quad,
\end{equation}
with $f^{n/A}$ constructed using the isospin symmetry.

In Figs.~\ref{fig:carbon1} and~\ref{fig:carbon2} we present the results for ${}^{12}$C.
Figure~\ref{fig:carbon1} shows the magnitude of the PDF $x\, f(x,Q)$ at $Q=2$~GeV on a log-scale. 
We observe that the separate fits are generally similar across the full $x$ kinematic range 
and that there is no abnormal behavior in either the low or high $x$ region. 
Additionally, this figure underscores the extent to which the PDFs are decreasing 
as we move to larger $x$,
especially for the gluon and sea-quarks $\{g, \bar{u}, \bar{d}, s{=}\bar{s}\}$; 
obtaining large event statistics in the high-$x$ region 
is challenging. 

\subsubsection{The Small \texorpdfstring{$x$}{x} Region} \label{sec:smallx}

In Fig.~\ref{fig:carbon2} we display the ratio of the fits compared to 
the nCTEQ15 nPDFs, we use here a log-linear scale to better illustrate the full $x$ range. 
Our first observation is that all of our new fits to the JLab data 
yield very comparable results in the small $x$ region. This is a reasonable 
result as the different corrections we are applying in the fits 
({\it e.g.}, HT, DEUT) all primarily impact the large $x$ region. 
We also observe that all these fits are uniformly below the nCTEQ15 PDFs
by ${\sim}6\%$.
Thus, it appears the fits are effectively hardening the high-$x$ PDFs
of select parton flavors (especially those of the $u$- and $d$-quark), while
somewhat suppressing PDFs in the small-$x$ region.
This behavior  suggests increased shadowing and merits further investigation. 

\subsubsection{The Large \texorpdfstring{$x$}{x} Region} \label{sec:largex}

From Fig.~\ref{fig:carbon2}
it is clear that the fits are 
hardening the nuclear PDFs at larger
$x$ to better
describe the new data sets. In particular, we see the up, down and gluon PDFs 
all are larger than nCTEQ15 while the sea quarks  
$\{\bar{u}, \bar{d}, s{=}\bar{s}\}$ are smaller.
For example, if we look in the region of $x{\sim}0.7$ we see 
the $\bar{u}$ and $\bar{d}$ PDFs are reduced by as much as ${\sim}10\%$ 
while the gluon is increased by ${\sim}5\%$
and the up and down PDFs are increased by ${\sim}10\%$.

Comparing the individual fits, we see a very clear pattern in the 
$\bar{u}$ and $\bar{d}$ PDFs. 
As we move from  the nCTEQ15 fit, to the  BASE, HT, DEUT, and finally 
the nCTEQHIX fit we see the $\bar{u}$ and $\bar{d}$ PDFs are uniformly decreasing in the large $x$ region. 
Recall that this ordering of the fits also shows a monotonically  decreasing value of $\chidof$, 
{\it c.f.},~Table~\ref{tab:chi2}. 

We also observe that the BASE and the HT fits are quite similar, as the addition of 
the HT corrections, 
that are substantially smaller for structure function ratios than for either the numerator or the denominator, see Fig.~\ref{fig:ht}, 
only show a slight improvement in the $\chidof$. 

Similarly, the DEUT and the \ncteqhix{} are also quite similar as these fits only 
differ by the inclusion of the HT corrections (for the \ncteqhix{} fit).
This difference is barely discernible in the figures 
as the curves are generally within a percent,
but  \ncteqhix{}  generates a 3\% improvement in $\chidof$ compared to DEUT.
These two fits also produce the largest downward shift of the  $\bar{u}$, $\bar{d}$ and strange PDFs
while yielding a smaller increase of the up and down PDFs. 
Recall that  DEUT and \hbox{\ncteqhix{}} have the lowest $\chidof$ values.

\subsubsection{Large  A Results} \label{sec:largea}

Having investigated the carbon results in detail, we now want to explore the 
larger $A$ values to see how the above effects scale to heavier nuclei. 
In Figures~\ref{fig:iron1}~and~\ref{fig:lead1} 
we display the PDF ratios (compared to nCTEQ15) 
for  iron and lead PDFs on a log-linear scale. 

The small $x$ behavior is quite similar to the carbon PDF result. 
At $x{=}10^{-3}$ all the PDFs are shifted downward by  ${\sim}6\%$
to  ${\sim}8\%$, and the shift of the individual fits is quite uniform. 

In the large $x$ region, the behavior is also similar to the carbon PDFs,
but the size of the shifts is slightly larger. 
For example, at $x{\sim}0.7$ the  $\bar{u}$ and $\bar{d}$ PDFs for carbon 
were shifted downward by ${\sim}10\%$ compared to nCTEQ15, 
while the  iron 
and lead are shifted by up to ${\sim}30\%$.
We again see that the gluon, up and down PDFs are increased at large $x$. 
Again comparing at $x{\sim}0.7$, while the up and down PDFs of carbon 
are increased by  ${\sim}10\%$,   the iron and lead are shifted by up to   ${\sim}20\%$.

Clearly the new data has significant impact on the PDFs in this kinematic region. 
We also observe that the impact of the individual corrections \{HT, DEUT\} 
are rather uniform, and we obtain the best fit using both the HT and DEUT corrections 
which give us the \ncteqhix{} fit. 
Therefore in the following, we will now focus on this fit.

\subsubsection{PDF Uncertainties} \label{sec:uncert}

Finally, we now compare the uncertainties of the 
nCTEQ15 and  \ncteqhix{} PDF sets as displayed in 
Figs.~\ref{fig:carbon1}--\ref{fig:lead1}.
These results should be interpreted carefully as we must be attentive to 
parameterization bias and other imposed constraints. For example,
our parameterization ties the strange PDF to the average $(\bar{u}+\bar{d})$ distribution, 
so this is not a fully independent distribution.%
    \footnote{Note, our recently released \ncteqwz{} nPDF analysis included
    LHC $W/Z$ production data which allowed to provide additional freedom
    (and constraints) for the strange PDF~\cite{Kusina:2020lyz}.} 

Examining Fig.~\ref{fig:carbon1} showing the ${}^{12}$C PDFs, 
both nCTEQ15 and \ncteqhix{} have comparably sized uncertainty bands,
with the exception of the gluon. The larger uncertainty band for the \ncteqhix{} 
gluon may be more representative of the true uncertainties as it can be difficult
to fully explore the loosely constrained gluon in the small $x$ region 
where minimization methods often encounter troublesome flat parameter directions. 
In this sense, the larger gluon uncertainty band of, for example, the EPPS16 nPDF set~\cite{Eskola:2016oht} may be 
more characteristic ({\it c.f.}, Fig.~\ref{fig:others1}).

Note that the nCTEQ15 fit includes data on single inclusive pion
production from the STAR and PHENIX experiments, which we do not
include in the \ncteqhix{} fit.  However, we have verified these experiments
are compatible with our
\ncteqhix{} fit, and they yield a  $\chidof{} \sim 0.5$.
The pion data can help constrain the nuclear gluon PDF, especially in the
mid- to lower-$x$ region,
and to some extent, the slight increase of the gluon uncertainty
of \ncteqhix{} can thus be attributed to this.

Turning to the PDF ratio plots of Figs.~\ref{fig:carbon2}--\ref{fig:lead1},
again these should be interpreted carefully, especially in the large $x$ region 
where we have a ratio of small numbers. 
We observe that both the up and down PDFs appear to have reduced uncertainty bands
in the larger $x$ region. This makes sense as the bulk of our new data 
is constraining these flavors in this kinematic region. 
The gluon uncertainty is increased compared to nCTEQ15, presumably for the 
same reasons discussed above. 
The $\bar{u}$ and $\bar{d}$ uncertainties are roughly comparable between the
nCTEQ15 and \ncteqhix{}, as we expect only indirect constraints on these flavors.

While the uncertainties of the deuteron structure functions displayed in Fig.~\ref{fig:deut}
are not used in the computation of the error bands, their impact can be assessed by 
comparing the fits without (BASE, HT) and with (DEUT, \ncteqhix{}) the deuteron corrections.
These differences lie within the uncertainty bands in all cases, except for 
the up and down nPDFs at intermediate $x$ values ($x{\sim}0.5$);
hence, our uncertainties in this region may be slightly underestimated.

Finally, the strange PDF also shows a reduced uncertainty band.
As mentioned in Sec.~\ref{sec:parms},
this distribution is tied to the  $(\bar{u}+\bar{d})$ sea quarks
and we have not introduced any free strange parameters.
For a more representative uncertainty estimate  of the strange PDF,
compare with the error bands of  \ncteqwz{} ({\it c.f.}, Ref.~\cite{Kusina:2020lyz}, Fig.~12).

\subsection{Parameter Scans} \label{sec:scans}

\begin{figure*}[ptb]
\begin{subfigure}
\centering
  \begin{subfigure}{a)}
\includegraphics[width=0.45\textwidth]{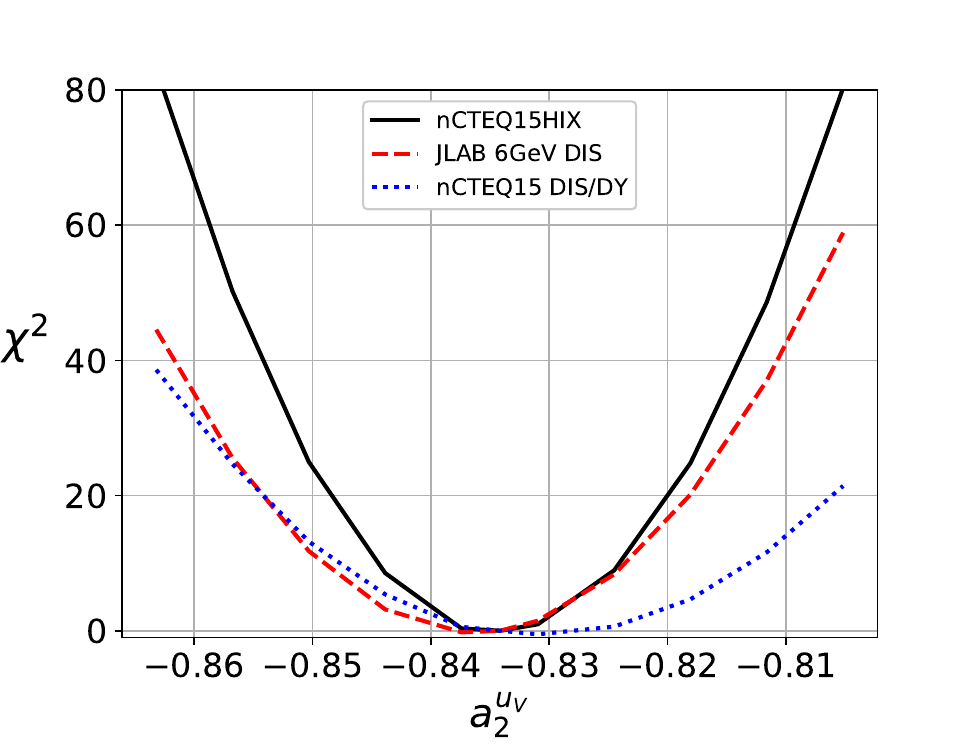} 
  \end{subfigure}
\hfil
  \begin{subfigure}{b)}
\includegraphics[width=0.45\textwidth]{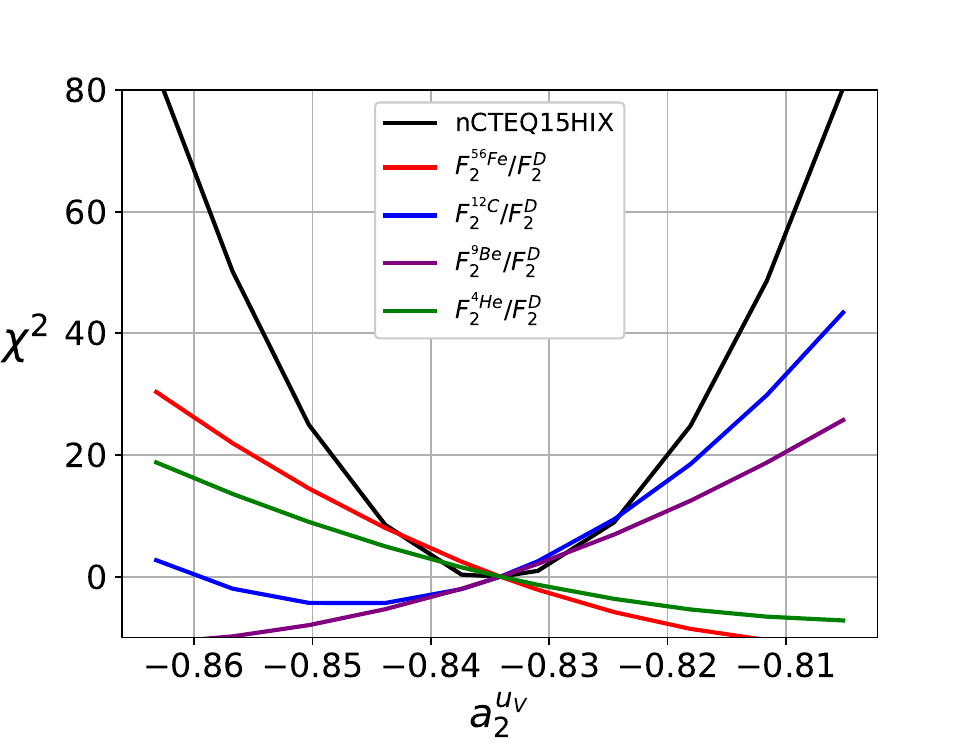} 
  \end{subfigure}
\caption{PDF parameter profiles of the the up-valence parameter $a_2^{u_v}$. 
Fig.~a)~We display the $\chi^2$ shift due to the original nCTEQ15 data set (blue dots),  
the new JLab data set (red dashes), and the total (black solid). 
Fig.~b)~We display the $\chi^2$ shift for the most significant projectile-target combinations constraining the $a_2^{u_v}$ parameter.
}
\label{fig:uvprofile}
\end{subfigure}
\vspace{0.5cm}
\begin{subfigure}
\centering
  \begin{subfigure}{a)}
\includegraphics[width=0.45\textwidth]{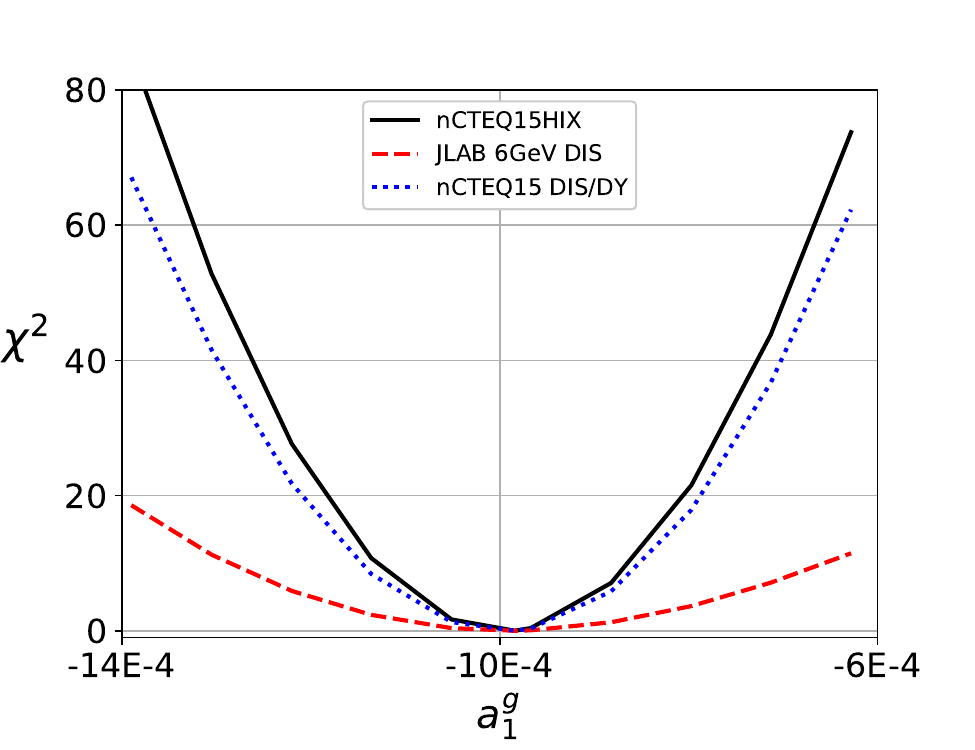} 
  \end{subfigure}
\hfil
  \begin{subfigure}{b)}
\includegraphics[width=0.45\textwidth]{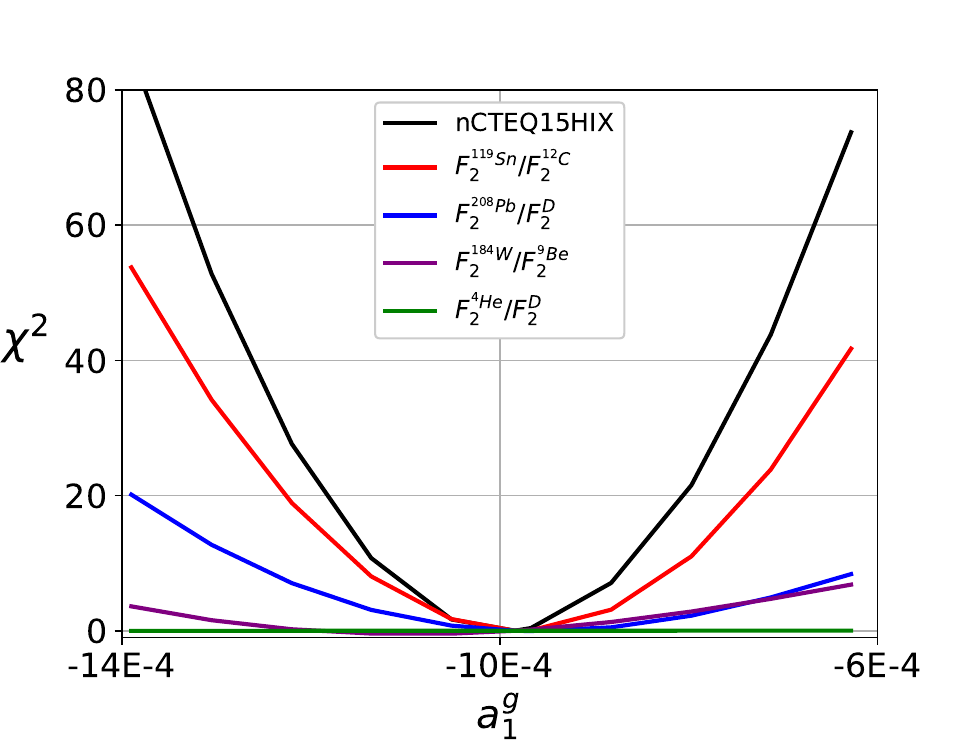} 
  \end{subfigure}
\caption{PDF parameter profiles of the the gluon parameter $a_1^{g}$. 
Fig.~a)~We display the $\chi^2$ shift due to the original nCTEQ15 data set (blue dots),
the new JLab data set (red dashes), and the total (black solid). 
Fig.~b)~We display the $\chi^2$ shift for the most significant projectile-target combinations constraining the $a_1^{g}$ parameter. 
}
\label{fig:gprofile}
\end{subfigure}
\end{figure*}

We have performed parameter scans for the 19 parameters of the \ncteqhix{} fit,
and selected results are displayed in Figs.~\ref{fig:uvprofile} and~\ref{fig:gprofile}.
As expected, the JLab data 
exerts strong constraints on the parameters of the up and down quarks, 
while the impact on the sea quarks and gluon is reduced. 
We also note that for the sea quarks (not shown) and gluon, the value of the parameters 
at the minimum $\chi^2$ is quite similar for both the nCTEQ15 set and the new JLab set.
However, for the up and down parameters, where the JLab data has more constraining power, 
we find in some cases ({\it e.g.}, $a_2^{u_v}$) the fit minimum is a compromise 
between the two data sets. 
This is a common feature in global PDF fits with large data sets. 
To gain additional insight on which data sets are driving this result
we now examine  Figs.~\ref{fig:uvprofile} and~\ref{fig:gprofile} in detail. 

In  Fig.~\ref{fig:uvprofile} we display the profile for the $a_2^{u_v}$ parameter.
The first plot, Fig.\ref{fig:uvprofile}-a), shows the total $\chi^2$ shift,
as well as the individual shifts due to the original nCTEQ15 data set and
the new JLab data set.
Clearly the new JLab data help constrain this parameter compared to the nCTEQ15 data alone. 
We also observe the original nCTEQ15 data prefers slightly larger values for the $a_2^{u_v}$ parameter
as compared to the JLab data; this  type of behavior is common for global PDF fits. 
In  Fig.~\ref{fig:uvprofile}-b) we display the $\chi^2$ shift for the most significant data sets constraining the $a_2^{u_v}$ parameter. We note the  data sets with the largest impact 
are the  ${}^{56}$Fe/D data  which is  pulling $a_2^{u_v}$ to higher values, and 
the ${}^{12}$C/D data which is pulling $a_2^{u_v}$ to lower values.

In  Fig.~\ref{fig:gprofile} we display the profile for the $a_1^{g}$ parameter.
Fig.\ref{fig:gprofile}-a) again shows the total $\chi^2$ shift,
as well as the individual shifts due to the original nCTEQ15 data set and
the new JLab data set.
We observe the new JLab data provides a weaker constraint on the gluon 
as compared to the valence quarks. 
We also note the preferred minima of both the original nCTEQ15 data 
and the JLab data coincide for the gluon parameter.
Turning to   Fig.~\ref{fig:gprofile}-b)  we see  the $\chi^2$ shift for the most significant data sets constraining the $a_1^{g}$ parameter. 
In contrast to the above case of $a_2^{u_v}$, 
we note the  data sets with the largest impact 
are   ${}^{119}$Sn/${}^{12}$C and ${}^{208}$Pb/D; in this case, 
the  ${}^{56}$Fe/D   data has less impact on this parameter.

\section{The Low-\textit{W} Kinematic Region}
\label{sec:noWcut}

The $W{>}1.7$ GeV cut on the data fitted in the \ncteqhix{} analysis discussed above was imposed to restrict the study to the DIS regime,
where we can reasonably assume QCD factorization to work, and avoid resonance excitations in electron-scattering events. However, it has
been shown that the characteristic $x$ dependence of nuclear structure-function ratios displayed, for example, in Fig.~\ref{fig:sample1a}
persists into the resonance region at low $W$ values. Specifically, Fig.~3 of Ref.~\cite{Arrington:2003nt} overlays data from the resonance
region ($1.1\! <\! W\! <\! 1.7$~GeV, $Q{\sim}2$~GeV) with $W\!>\!1.7$ GeV DIS data from the SLAC E139~\cite{Gomez:1993ri}, SLAC E87~\cite{Bodek:1983qn},
and BCDMS~\cite{Benvenuti:1987az} experiments, and finds that ``the agreement of the resonance region data with the DIS measurement [...] is
quite striking.''
This is likely a manifestation of the quark-hadron duality phenomenon, abetted in the case of nuclear scattering by
Fermi motion of the bound nucleons, which results in a further, effective averaging of the nucleon structure functions over multiple resonances; together,
these dynamics may permit a description of nuclear structure functions in terms of partonic degrees-of-freedom, even in kinematic
regions where resonance excitation is the dominant effect.
It is therefore interesting to remove the $W$ cut, and explore to what extent the resonance region data can be described in terms of nPDFs.

\subsection{nPDFs at large \textit{x}
}
\label{sec:smearing}

\begin{figure}[tb]
\includegraphics[width=0.45\textwidth]{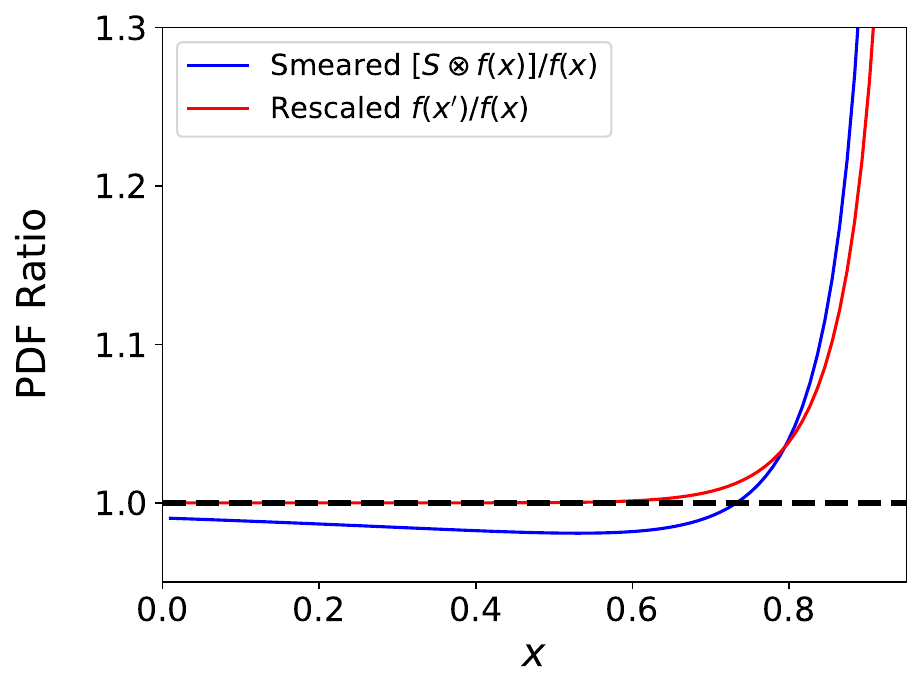}
\caption{We display the ratio of the modified (smeared/rescaled) PDF compared to the 
unmodified PDF for both the Gaussian smearing and the $x$-rescaling. 
For the Gaussian smearing (blue), we display $S\otimes f(x)$ over $f(x)$ with 
$\Delta=0.05$, and $\delta=-0.007$ (which resembles an $A=12$ nucleus).
For the $x$-rescaling (red), we display $f(x'_A)/f(x)$
for  $\varepsilon=0.02$, $\kappa=10$, and  $A=12$.
While there is a small relative difference at intermediate to small $x$ values,
the rise at large $x$ is quite similar for both methods. 
}
\label{fig:rescale2}
\end{figure}
\begin{figure}[tb]
\includegraphics[width=0.45\textwidth]{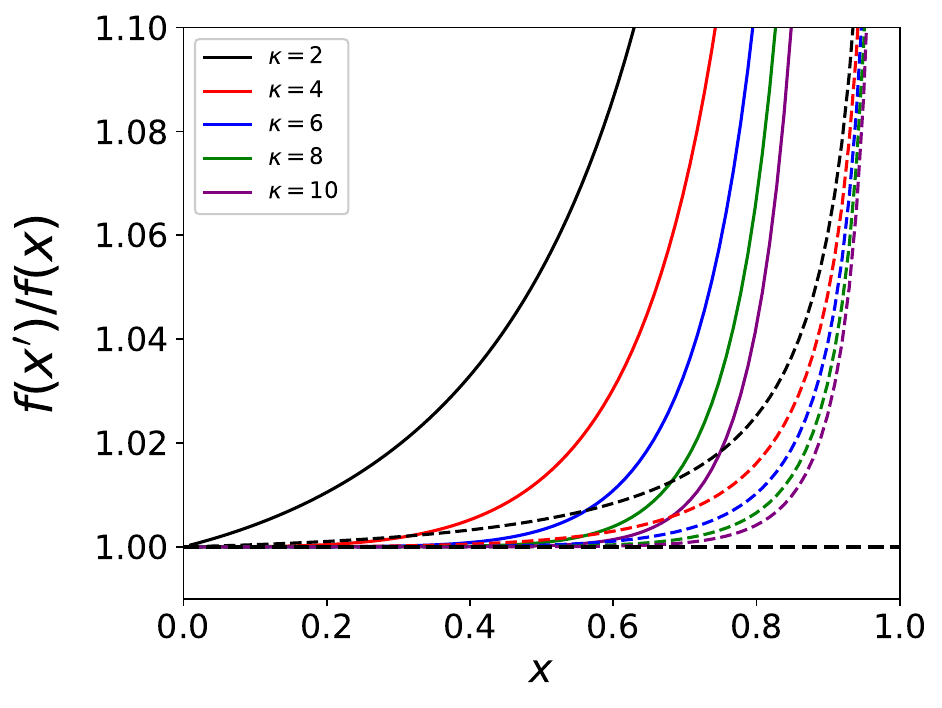}
\caption{We show the impact of the $x$-rescaling of Eq.~(\ref{eq:rescale})
for varying choices of $\varepsilon$ and $\kappa$
for a ``toy'' PDF, $f(x)\sim {(1-x)^3/x^{1.5}}$.
The plot shows the ratio $f(x'_A)/f(x)$ for 
2 groups of 5 curves for $A=208$.
The largest 5 curves (solid) are for $\varepsilon=0.01$ 
with exponent $\kappa=\{2,4,6,8,10\}$ indicated by the colors
\{black, red, blue, green, purple\}, respectively. 
The lower 5 curves (dashed) are for $\varepsilon=0.001$
with the same range of $\kappa$ values.
}
\label{fig:rescale1}
\end{figure}

An important difference between nuclear PDFs and proton PDFs is that, in principle, the momentum fraction $x$ of a parton with respect to the
average nucleon momentum in a nucleus can range from $[0,A]$ whereas  that of a proton is confined to lie in the range $[0,1]$. 
Phenomenologically, at the photon energies under consideration, nuclear partons belong to bound nucleons and share a relative fraction $x_N \in [0,1]$ of their momentum. As the nucleons interact among themselves, they exchange momentum; thus, their partons can receive a boost in the photon's direction, causing their momentum fraction $x$ to shift and exceed 1. Since the nucleon-nucleon interaction is on average soft, the shifts are moderate and the nuclear parton distributions are still predominantly within $x\sim[0,1]$ with only a small tail exceeding 1.

This effect is currently not captured in the nPDF parametrization adopted in this paper, see Sect.~\ref{sec:parms}, which is confined to $x \in [0,1]$ and needs to be suitably extended. In fact, a full description of nPDFs in the $x \in [0,A]$ range would also entail a generalization of DGLAP evolution equations and careful consideration of the sum rules \cite{Freese:2015ebu}.\footnote{Note that restricting our formalism to $x \in [0,1]$, i.e., assuming $f_i^A(x,Q)=0$ for $x>1$ for all parton flavors $i$ as done in all current nPDF analyses, is 
self-consistent under scale-evolution and preserves the parton sum rules.}
In this paper we limit ourselves to an exploratory study and demonstrate, as a proof of principle, that a simple generalization of the adopted parametrization can lead to a reasonable description of electron-nucleon collisions in terms of nuclear PDFs even in the resonance region. We will leave more detailed studies for future work.

\subsubsection{Convolution kernel}

The parton momentum shift caused by nucleon-nucleon interactions can be effectively described by a convolution of the nucleon-level PDF $f$ with a smearing function $S_A$:
\begin{align}
    f_A(x,Q) = \int_x^A \frac{dy}{y} S_A(y,Q) f(x/y,Q) \ ,
\label{eq:f_smearing}
\end{align}
where $S_A$ represents the probability that the parton $f$ belongs to a nucleon of momentum fraction $y$ compared to the average nucleon momentum. Such  representation is, in fact, grounded on the so-called Weak Binding Approximation to the calculation of nuclear structure functions, where $S_A$ can be obtained starting from the nucleon wave function in the corresponding nucleus~\hbox{\cite{Kulagin:1989mu,Kulagin:1994fz,Kulagin:1997vv,Kulagin:2004ie,Kulagin:2007ju,Kulagin:2010gd}}. 
In the Weak Binding Approximation, the smearing function is in general scale-dependent at finite $Q$ values. Therefore $f_A$ does not satisfy the DGLAP evolution equations except at $Q \to \infty$ where $S_A$ becomes scale independent, or at small $x$ where the smearing effect is marginal. In the context of nPDF fits, however, we rather regard the convolution of Eq.~\eqref{eq:f_smearing} as a generalization of the nPDF parametrization discussed in Sect.~\ref{sec:parms}, and would only apply it at the initial scale~$Q_0$.

At $x$ values not exceeding 1, the nucleon-nucleon interaction is soft and the smearing function can be approximated by a normalized Gaussian, 
\begin{equation}
    S_A(y) = \frac{1}{\sqrt{2\pi \Delta_A}} 
    \exp{\left\{ \frac{-[y-(1-\delta_A)]^2}{2 \Delta_A^2} \right\} }
    \quad ,
\label{eq:Gaussian_smear_function}
\end{equation}
where the effect of nucleon-nucleon interactions is described by the Gaussian width $\Delta_A$. Furthermore, the nucleon binding causes a (small) shift of the Gaussian peak towards smaller $x$ values,
and this is controlled by the $\delta_A$ term. 
Note that the larger the nucleus, the more the nucleons that interact with each other (hence a larger $\Delta_A$) and the larger the binding energy (hence a larger $\delta_A$).

The net result of this convolution is to deplete the partons in the intermediate $x$ region, and enriching the large $x\gtrsim 1$ region. 
Hence the nuclear ratio $F_2^A/F_2^B$ with $A>B$ will also display a dip below unity at intermediate $x$ values (the EMC region) and a relatively steep rise above unity in the $x \to 1$ Fermi motion region, as seen experimentally. With a suitable choice or fit of the $\Delta_A$ and $\delta_A$ parameters one can therefore expect to quantitatively describe the observed nuclear effects
in the Fermi motion region\footnote{In principle, by generalizing the Gaussian \textit{ansatz} in Eq.~\eqref{eq:Gaussian_smear_function} for the smearing function and including, for example, power law tails \cite{CiofidegliAtti:1995qe}, the convolution formalism of Eq.~\eqref{eq:f_smearing} can also be used to extend the analysis into the $x{>}1$ region. We leave this for future work.}.

As an example, in Fig.~\ref{fig:rescale2} we show the ratio of convoluted to unmodified PDF for a toy model $f(x) = x^{-1.5} (1-x)^3$, which reproduces the main features of the up and down quark PDFs. The parameters have been chosen to approximate the behavior of the Carbon structure function data. 

\subsubsection{\textit{x}-rescaling}

While the smearing approach has a close connection to the underlying nuclear dynamics and guarantees that the PDF sum rules are satisfied, the convolution is computationally expensive. Hence, in this exploratory study we will try 
a simple remapping of the parametrization considered so far, by an A-dependent shift of the $x$ variable.

We want this remapping to approximate the rise of  $F_2^A /F_2^p$  in the large $x$ region
without distorting the phenomenology in the intermediate to small $x$ region. 
To ensure these conditions are satisfied, the remapping should 
(i)~only impact the very large $x$ region ($x{\sim}1$),
and (ii)~ensure the various momentum sum rules are satisfied within uncertainties. 

The specific remapping we shall use is
\begin{align}
    f_A(x,Q) \longrightarrow f_A(x'_A,Q)    
\end{align}
where $f_A$ are the PDFs determined in the nCTEQ15HIX fit, and
\begin{equation}
    x'_A = 
    x-   \varepsilon \, x^\kappa \log_{10}A
    \ .
    \label{eq:rescale}
\end{equation}
As the PDFs are typically decreasing functions of the parton's fractional momentum, the negative shift ensures that the transformed function is larger than the unmodified one, and non-vanishing as $x \to 1$. 
The $\varepsilon$ parameter controls the overall size of the rescaling effect, and 
the $x^\kappa$ factor ensures that only the large-$x$ region is modified.
The $\log_{10} A$ term ensures an increasing modification across the full range of nuclear $A$ values  from proton $(A=1)$ to lead $(A=208)$.

In Fig.~\ref{fig:rescale1}, we show the impact of the rescaling on the ratio 
$f_A/f_p$ for $A=208$ and a selection of $\varepsilon$ and $\kappa$ values.
For simplicity, we have used here the same $f(x) =  {(1-x)^{3}/x^{1.5}}$ toy PDF considered in the nuclear smearing discussion. 
Clearly, the remapping does not preserve the PDF normalization within the $x\in[0,1]$ range, and therefore breaks the PDF sum rules. However, in all cases the breaking amounts to a small increase
of the momentum sum by less than 1\%, which is well within the uncertainty of the nPDFs.

\subsection{nPDFs in the \texorpdfstring{$x{\to}1$}{x->1} Limit} \label{sec:nwc}

\begin{figure*}[t]
\includegraphics[width=0.95\textwidth]{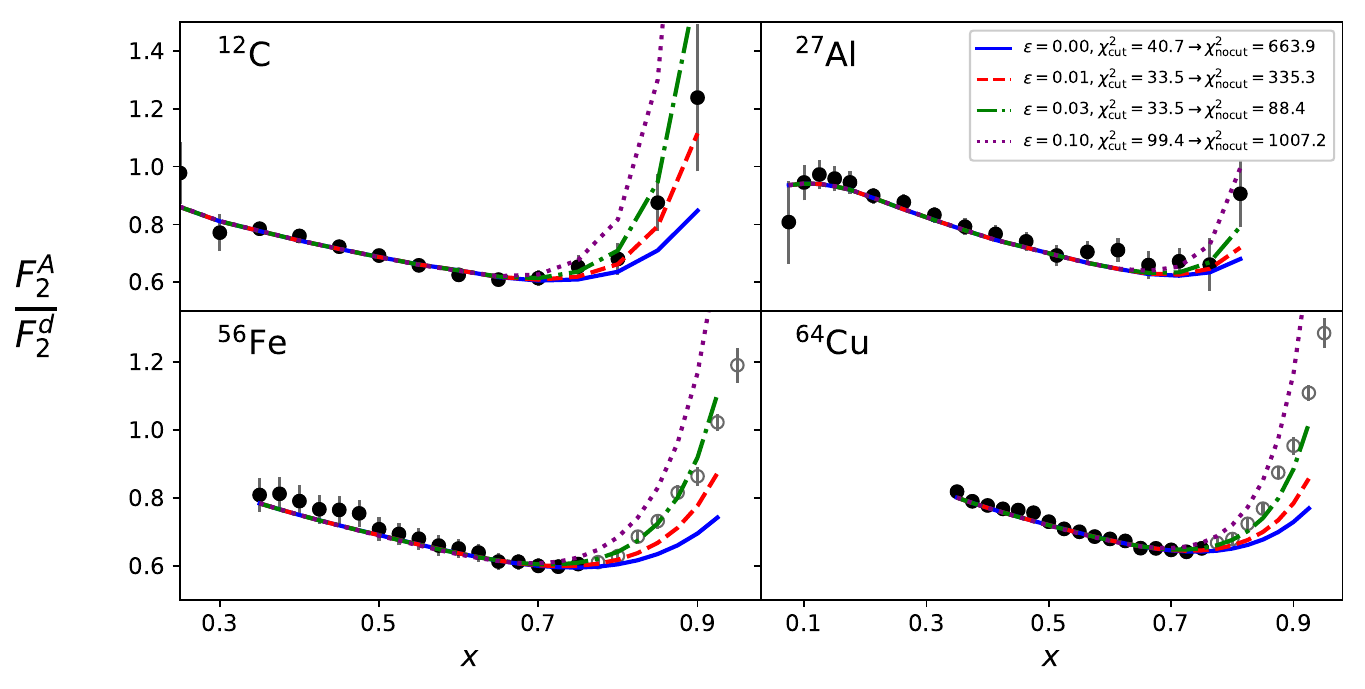}
\caption{We show $F_2^A/F_2^D$ for selected data sets 
in the large $x$ region. The different curves show the impact
of the variations of $\epsilon=\{0, 0.01,0.03,0.1 \}$ 
(solid blue, dashed red, dot-dashed green, dotted purple)
with $\kappa=10$ in the rescaling of Eq.~(\ref{eq:rescale}).
The combined $\chi^2$ values  are displayed in 
the legend, both with and without the $W>1.7$~GeV cut. 
For reference, the bottom curve (blue, solid) in each plot corresponds to $\epsilon=0$ which is the \ncteqhix{} result. 
Points which are excluded from the \ncteqhix{} fit by the $W>1.7$~GeV cut are indicated with a 
hollow (gray) symbol; 
this is   evident for the Fe and Cu data sets.
For reference, experiment 
Fe/D is ID=5131,
Al/D is 5134,   
Cu/D is 9984,  and
C/D  is 9990.   
}
\label{fig:nowcut}
\end{figure*}

Overall, the remapping accomplishes our stated goals, and can be used to investigate how well it can describe the data at the highest $x$, that was excluded from our \ncteqhix{} fit due the $W>1.7$ GeV cut.

To simplify the discussion, we fix the small-$x$ suppression parameter to $k=10$ and will study the effect of varying the $\epsilon$ parameter. One can appreciate that this choice is reasonable by looking at Fig.~\ref{fig:rescale2}, where the $A=12$ remapped $f_A/f_p$, plotted in red, is compared with the blue convolution model discussed earlier. By design the remapping only deforms the PDF at large $x$, and does not reproduce by itself the intermediate-$x$ dip displayed by the convolution model. This however is small and can be easily taken care of in a fit by the standard nPDF parameters discussed in Sect.~\ref{sec:parms}. On the contrary, the large-$x$ steep rise is pretty well approximated by the remapped PDF. We have furthermore verified that $k=10$ is also a good choice for other nuclear targets, given the assumed logarithmic $A$ dependence of the rescaled $x'_A$ momentum fraction. 

We can now investigate the impact of rescaling the \ncteqhix{} nPDFs
and compare these to the high-$x$ data. 
In Fig.~\ref{fig:nowcut} we display a selection of data sets that have data points in the $x\to1$ region, and compare them to calculations performed with shifted nPDFs for fixed $\kappa=10$ and a representative choice of $\epsilon$ parameters in the $[0,0.1]$ range. Note that in the ${}^{56}$Fe and ${}^{64}$Cu panels, the empty symbols represent data points at $W<1.7$~GeV, that were excluded from the nPDF fit. 
We also display the computed $\chi^2$ values in the plot legend. 
These are computed both for the points that satisfy the $W>1.7$~GeV cut  (labeled $\chi^2_\text{cut}$), and for all the points (labeled $\chi^2_\text{nocut}$). Note that for these $\chi^2$ calculations we have simply applied the $x$ rescaling when computing $F_2(x,Q)$ using the \ncteqhix{} nPDFs; this is not a separate fit. 

If we focus on the top two panels of  Fig.~\ref{fig:nowcut},
${}^{12}$C and ${}^{27}$Al, we see all these data points satisfy 
the $W>1.7$~GeV cut.
Turning our attention to the bottom two panels of  Fig.~\ref{fig:nowcut},
${}^{56}$Fe and ${}^{64}$Cu, some  data points do not satisfy 
the $W>1.7$~GeV cut, and these points give rise to 
the differing $\chi^2_\text{cut}$ and  $\chi^2_\text{nocut}$  values.

Overall, we observe there is a small but significant variation in  $\chi^2_\text{cut}$, 
and a dramatic variation in $\chi_\text{nocut}^2$. In both cases, the fit shows preference for $\epsilon \approx 0.03$. With this parameter, we can describe very well the turnover of the structure function ratio at $x\sim0.7$, and the subsequent large-$x$ rise over the whole A range. Overall, we obtain a very reasonable description of the large-$x$ data in the resonance region.

\subsection{Extending nPDF fits to low  \texorpdfstring{$W$}{W}}
\label{sec:recap}

We have found that the $x$ rescaling provides a simple means of mimicking the behavior of the nuclear PDF in the $x \to 1$ region. 
While we have used it here \textit{post-facto}, it can also be effectively utilized to generalize the nPDF parametrization discussed in Sect~\ref{sec:parms} with only 2 additional free parameters. As the $x$ rescaling already increases the $F_2^A/F_2^D$ ratio in the Fermi motion region by about the right amount compared to the baseline \ncteqhix{} nPDFs, the remaining parameters will more easily adjust to the data, and we can expect an improvement in the fit quality in the high $x$, low $W$ region. 

Furthermore, this analysis suggests that it may be possible to expand  the kinematic reach of the fit in the $\{x,Q\}$-space to include data in the low $W<1.7$ GeV resonance region. This is an interesting possibility, and we reserve the details for a future study.

\section{Comparisons with other \texorpdfstring{\MakeLowercase{n}}{n}PDFs}
\label{sec:compare}
\begin{figure*}[tbhp]
\includegraphics[width=0.98\textwidth]{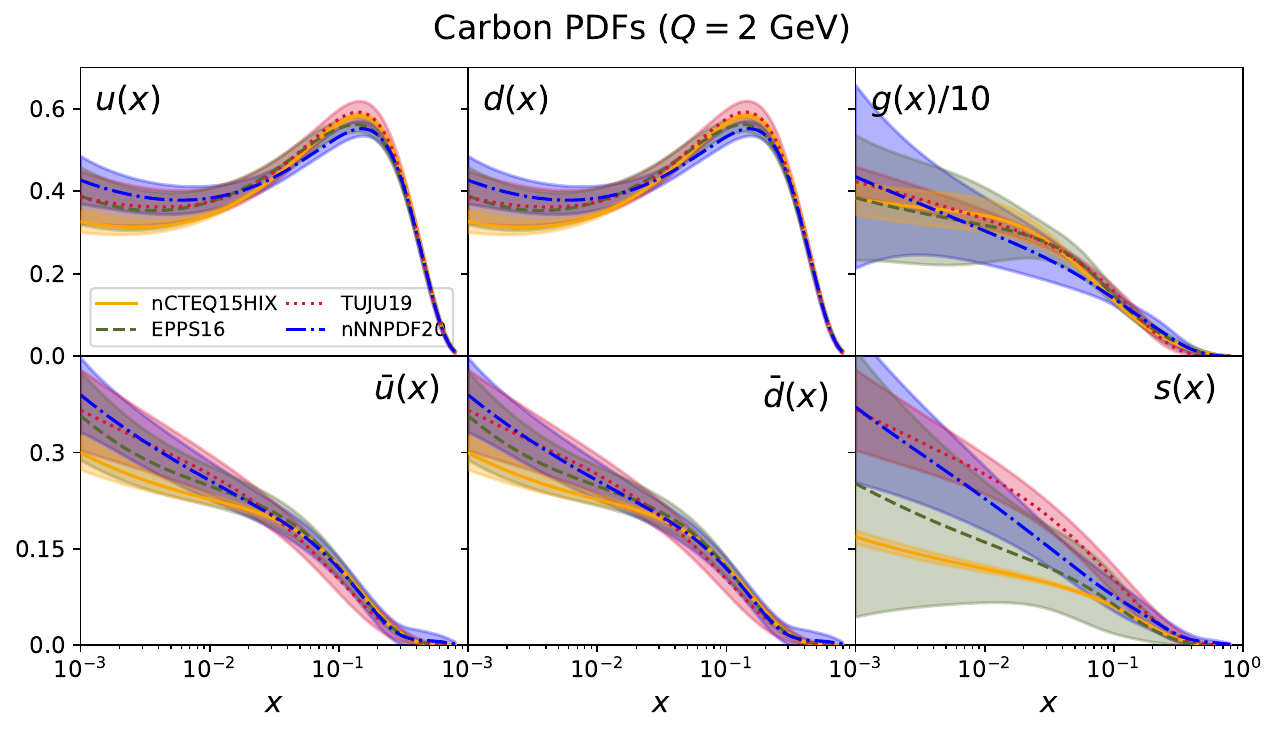}
\caption{We compare our \ncteqhix{} results to other nPDF sets from the literature
including EPPS16~\cite{Eskola:2016oht}, nNNPDF2.0~\cite{AbdulKhalek:2019mzd},
and TUJU19~\cite{Walt:2019slu}. 
We plot $xf(x,Q)$ for carbon ${}^{12}$C at $Q=2$~GeV  
on a log scale. 
}\label{fig:others1}
\end{figure*}
\begin{figure*}[tbhp]
\includegraphics[width=0.98\textwidth]{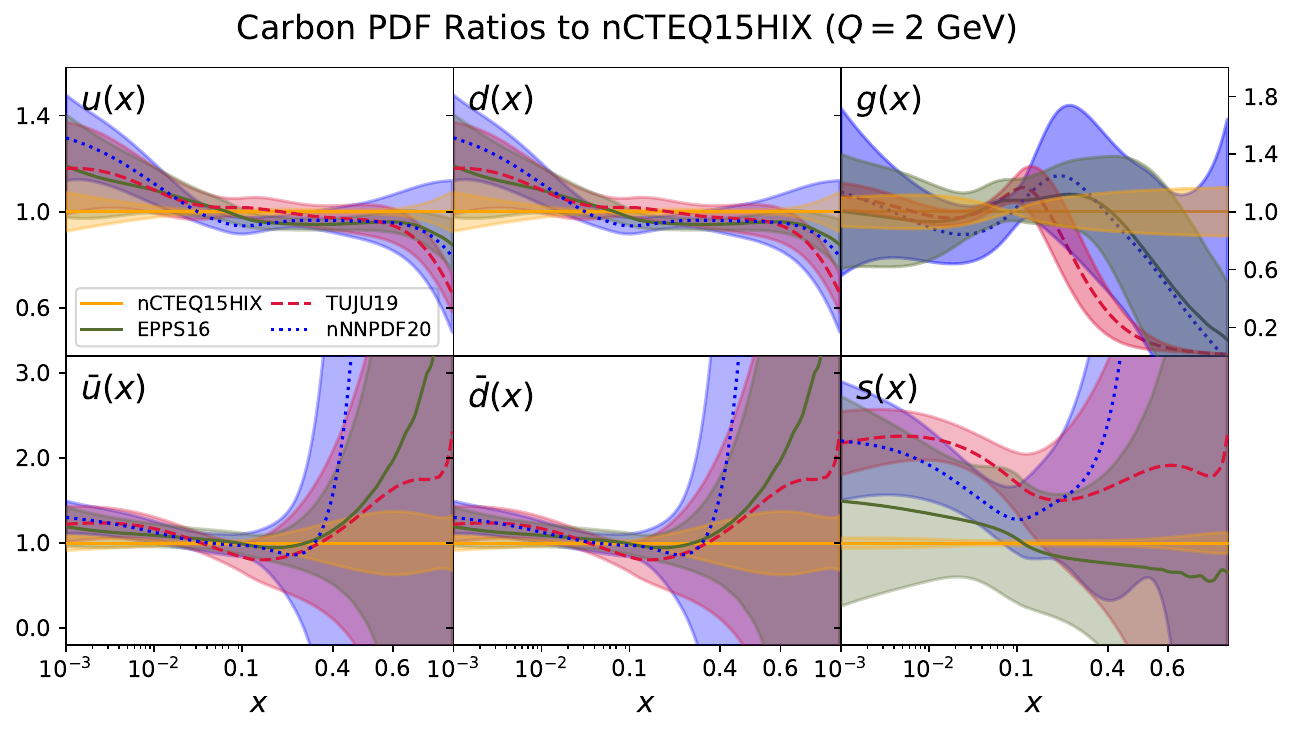}
\caption{We compare our \ncteqhix{} results to other nPDF sets from the literature
including EPPS16~\cite{Eskola:2016oht}, nNNPDF2.0~\cite{AbdulKhalek:2019mzd},
and TUJU19~\cite{Walt:2019slu}. 
We plot the nPDF ratio for carbon ${}^{12}$C at $Q=2$~GeV  
compared to \ncteqhix{} 
on a log-linear scale.
\\
}
\label{fig:others2}
\end{figure*}

In this section we compare our new \ncteqhix{} nPDFs with other results from the literature. 
In particular, in Figs.~\ref{fig:others1}  and~\ref{fig:others2} we compare with nPDFs from
EPPS16~\cite{Eskola:2016oht},
nNNPDF2.0~\cite{AbdulKhalek:2019mzd},
and TUJU19~\cite{Walt:2019slu} analyses.
We find the results are generally compatible within the uncertainty bands. 

Examining the ratios of Fig.~\ref{fig:others2} we note some features similar to the comparison 
between nCTEQ15 and \ncteqhix{} of Fig.~\ref{fig:carbon2}.
Specifically, for the up, down, and gluon PDFs,
\ncteqhix{} is generally above the other PDFs in the high $x$ region. 
Correspondingly,  for the anti-up and anti-down PDFs,
\ncteqhix{} is generally below the other PDFs in the high $x$ region. 
This is similar to the pattern observed in Fig.~\ref{fig:carbon2},
although these difference are well within the uncertainty bands. 
It will be of interest to follow these differences as the uncertainty bands are 
reduced in future analyses. 

We do not display a comparison with our recent nPDFs \ncteqwz{}~\cite{Kusina:2020lyz} where
the nCTEQ15 analysis was supplemented by the available $W/Z$ boson production data from the LHC.
This is because here  we concentrate on the high-$x$ region of quark distributions, 
and the $W/Z$ data were mostly relevant for the strange and gluon distributions at lower $x$
values. Hence, they had minimal impact on the high-$x$ behavior of the nPDFs.
Therefore, the high $x$ behavior of the \ncteqwz{} PDFs was quite similar to the   original nCTEQ15 nPDFs, and this  was already presented in Figs.~\ref{fig:carbon1}-\ref{fig:lead1}.

\section{Conclusion}
\label{sec:conc}
PDFs encode the dynamics of the strong interaction and provide a crucial link between experimental measurements and theoretical models. Thus, the updated nCTEQ15HIX nPDFs presented here will contribute to increased precision of experimental analyses, which can thereby yield further insights into the QCD.
The novel aspects of this study consisted of several elements.

The newly-included JLab data provide important additional constraints on the nuclear PDFs
in the high-$x$ and low-$Q$ regime. 
We have explored this region with the \ncteqhix{} set
by relaxing our kinematic cuts to $Q{>}1.3$~GeV and $W{>}1.7$~GeV. 

In the high-$x$ region, there are various theoretical corrections 
which must be taken into account, especially for the $Q\! \sim\!$ few GeV
region now included with our relaxed kinematic cuts.
Among these, the Higher-Twist (HT) modifications discussed in Sec.~\ref{sec:ht} reduce $\chidof$ by ${\sim}3\%$,
while the Deuteron (DEUT) modifications of Sec.~\ref{sec:deut} reduce $\chidof$ by ${\sim}10\%$; ultimately,
the combination of both (which was used in \ncteqhix{}) reduces $\chidof$ by ${\sim}15\%$.

We also investigated the low-$W$ kinematic region including the
impact of extending the nuclear PDFs 
into the region  of very large $x$.
We used 
a $x$-rescaling technique that mimics the parton momentum shift to higher $x$ induced by the nucleon's Fermi motion.
We find that the $x$-rescaling 
provides a simple (albeit approximate) means 
to extend the nPDF parametrization to larger $x$ values than currently included in our main fit, and to address data falling within the resonance region.
Calculations based on a simple two-parameter $x$ rescaling were indeed able to improve the $\chidof$ for data at large-$x$ not only 
inside our current cuts, but -- crucially and more markedly -- also for those data that are currently outside and do not contribute to the nPDF fits.
This preliminary analysis suggests it may be possible to further 
reduce our $W$ cut, extend the fits into the resonance region,
and further improve the precision of the nPDFs at large momentum fraction $x$.

The structure function ratio  \mbox{$(F_2^A/F_2^D) {\cdot} (F_2^D/F_2^p)_{CJ}$} 
(Fig.~\ref{fig:sample1b}) displays the characteristic EMC shape, and 
the qualitative $x$ dependence of the nuclear ratio in the EMC Region was
found to be consistent for the range of $A$ considered in this analysis.

Our parameter scans (Figs.~\ref{fig:uvprofile}-\ref{fig:gprofile}) 
show that while the new data provide substantive constraints on the up-valence and 
down-valence distributions, there are nonetheless some tensions between individual data sets. 
This behavior is common for global PDF fits, but some of the details,
such as the tension between the Fe and C data sets (Fig.~\ref{fig:uvprofile})
may warrant additional investigation.

A general feature of the new \ncteqhix{} fit is a relative
hardening of the nuclear PDFs involving shifts of select distributions toward
higher $x$, an effect realized in the enhancement of the up-valence, down-valence, and gluon distributions, 
and corresponding depletion of the sea-quark PDFs, $\{ \bar{u}, \bar{d}, \bar{s} \}$.
This pattern is apparent when comparing \ncteqhix{}  
to nCTEQ15 (Figs.~\ref{fig:carbon1}--\ref{fig:lead1}).
Additionally, we find similar behavior when comparing \ncteqhix{} 
to other nPDFs from the literature including EPPS16, nNNPDF2.0, and TUJU19
(Figs.~\ref{fig:others2}). 
While these variations are within the nPDF uncertainty bands, 
it will be interesting to see to what extent this behavior persists with improved precision.

As PDF global analyses represent a primary computational tool available to describe hadronic interactions and structure in the context of QCD,
the nCTEQ15HIX nPDFs will enable a new level of precision in the exploration of nuclear dynamics and particle phenomenology. In a broader sense,
these tools can validate features of the standard model to next-generation precision, aid in the search for discrepancies which may signal undiscovered phenomena, 
and thereby yield deeper insights into the  QCD theory and the structure of hadronic matter.

\section*{Acknowledgments}
We are grateful to 
C.~Leger,
S.~Li,
W.~Melnitchouk, 
P.~Nadolsky, 
J.~Owens,
R.~Ruiz,
and
N.~Sato 
for valuable discussion.
This work was supported by the U.S.~Department of Energy under Grant No.~DE-SC0010129. 
T.H.~acknowledges support from an EIC Center Fellowship. 
The work of T.J.~was supported by the DFG under grant  396021762{--}TRR257. 
Work in M\"unster is supported by the DFG through the Research
 Training Network 2149 ``Strong and weak interactions - from hadrons to
 dark matter'' and through Project-ID 273811115{--}SFB 1225 ``ISOQUANT.''
A.K.~is grateful for the support of Narodowe Centrum Nauki under grant no.\ 2019/34/E/ST2/00186.
The work of A.A. and C.K.E.~was supported by the  U.S.\ Department of Energy contract DE-AC05-06OR23177, under which Jefferson Science Associates LLC manages and operates Jefferson Lab. A.A. was furthermore supported by DOE contract DE-SC0008791. J.G.M. has been supported by Fermi Research Alliance, LLC under Contract No. DE-AC02-07CH11359 with the U.S. Department of Energy, Office of Science, Office of High Energy Physics.
The work of I.S.\ was supported by the French CNRS via the IN2P3 project GLUE@NLO.

\appendix  %
\section{Data Sets Used in Fit}
\label{sec:datatabs}

Table~\ref{tab:jlab} lists the JLab DIS data of the form $F_2^A/F_2^D$,
Table~\ref{tab:exp1} lists the nCTEQ15 DIS data of the form $F_2^A/F_2^D$,
Table~\ref{tab:exp2} lists the nCTEQ15 DIS data of the form $F_2^A/F_2^{A'}$,
and
Table~\ref{tab:exp3} lists the nCTEQ15 DY data of the form $\sigma_{DY}^{pA}/\sigma_{DY}^{pA'}$.

In Table~\ref{tab:jlab}
we indicate the total number of data in the sets,  and the remaining number after the  
\hbox{\ncteqhix{}} cuts. 
In Tables~\ref{tab:exp1}, \ref{tab:exp2} and~\ref{tab:exp3} 
we indicate the total number of data in the sets,  and the remaining number after the 
nCTEQ15 kinematic cuts (first) as well as the \hbox{\ncteqhix{}} cuts (second). 

\vspace{0.5cm}
{\bf NMC $F_2^D$:} 
The NMC $F_2^D$ data set ID=5160 of Table~\ref{tab:exp1} is the one case which is not a ratio of structure functions.
It serves as a cross-check of how well we describe absolute structure
functions in the deuteron for which we apply a dedicated correction,
instead of fitting as is done for heavier nuclei.
Recall from Sec.~\ref{sec:parms} that our proton parameters $p_k$ are fixed,
and for the special case of the deuteron ($A=2$) we construct this using isospin symmetry.
We then apply the  deuteron structure function modifications  
described in Sec.~\ref{sec:deut}
for the DEUT and \ncteqhix{} fits, and find this improves  $\chidof$ for  $F_2^D$ by ${\sim}1\%$.

\vspace{0.5cm}
{\bf Data Files:} 
We have provided a full list of the data used in this fit in the text file \hbox{``fitted\_data.txt''}
which is included in the arXiv version of this 
document.\footnote{%
See Supplemental Material at 
\url{http://link.aps.org/supplemental/10.1103/PhysRevD.103.114015} for an ASCII
file containing the data used in this fit.
} %

\begin{table}[!hbp]
\begin{tabular}{|c|c|c|c|c|c|}
\hline 
{\footnotesize{}$\mathbf{F_{2}^{A}/F_{2}^{D}:}$}  &  &  &  &  & {\small{}\#data}\tabularnewline
{\footnotesize{Observable}}  & {\small{}Experiment} & {\small{}ID} &  Ref.  & {\small{}\# data} & {\small{}after cuts}\tabularnewline
\hline 
\hline 
{\small{}$^{208}$Pb/D} & {\small{}CLAS} & {\small{}9976} & \cite{Schmookler:2019nvf} & {\small{}25} & 24\tabularnewline
\hline 
{\small{}$^{56}$Fe/D} & {\small{}CLAS} & {\small{}9977} & \cite{Schmookler:2019nvf} & {\small{}25} & 24\tabularnewline
\hline 
{\small{}$^{27}$Al/D} & {\small{}CLAS} & {\small{}9978} & \cite{Schmookler:2019nvf} & {\small{}25} & 24\tabularnewline
\hline 
{\small{}$^{12}$C/D} & {\small{}CLAS} & {\small{}9979} & \cite{Schmookler:2019nvf} & {\small{}25} & 24\tabularnewline
\hline 
\multirow{2}{*}{{\small{}$^{4}$He/D}} & \multirow{2}{*}{{\small{}Hall C}} & {\small{}9980} & \cite{Seely:2009gt} & {\small{}25} & 17\tabularnewline
\cline{3-6} \cline{4-6} \cline{5-6} \cline{6-6} 
 &  & {\small{}9981} & \cite{Seely:2009gt} & {\small{}26} & 16\tabularnewline
\hline 
\multirow{2}{*}{{\small{}$^{3}$He/D}} & \multirow{2}{*}{{\small{}Hall C}} & {\small{}9982} & \cite{Seely:2009gt} & {\small{}25} & 17\tabularnewline
\cline{3-6} \cline{4-6} \cline{5-6} \cline{6-6} 
 &  & {\small{}9983} & \cite{Seely:2009gt} & {\small{}26} & 16\tabularnewline
\hline 
\multirow{2}{*}{{\small{}$^{64}$Cu/D}} & \multirow{2}{*}{{\small{}Hall C}} & {\small{}9984} & \cite{Seely:2009gt} & {\small{}25} & 17\tabularnewline
\cline{3-6} \cline{4-6} \cline{5-6} \cline{6-6} 
 &  & {\small{}9985} & \cite{Seely:2009gt} & {\small{}26} & 16\tabularnewline
\hline 
\multirow{2}{*}{{\small{}$^{9}$Be/D}} & \multirow{2}{*}{{\small{}Hall C}} & {\small{}9986} & \cite{Seely:2009gt} & {\small{}25} & 17\tabularnewline
\cline{3-6} \cline{4-6} \cline{5-6} \cline{6-6} 
 &  & {\small{}9987} & \cite{Seely:2009gt} & {\small{}26} & 16\tabularnewline
\hline 
\multirow{2}{*}{{\small{}$^{197}$Au/D}} & \multirow{2}{*}{{\small{}Hall C}} & {\small{}9988} & \cite{Seely:2009gt} & {\small{}24} & 17\tabularnewline
\cline{3-6} \cline{4-6} \cline{5-6} \cline{6-6} 
 &  & {\small{}9989} & \cite{Seely:2009gt} & {\small{}26} & 16\tabularnewline
\hline 
\multirow{10}{*}{{\small{}$^{12}$C/D}} & \multirow{10}{*}{{\small{}Hall C}} & {\small{}9990} & \cite{Seely:2009gt} & {\small{}25} & 17\tabularnewline
\cline{3-6} \cline{4-6} \cline{5-6} \cline{6-6} 
 &  & {\small{}9991} & \cite{Seely:2009gt} & {\small{}17} & 7\tabularnewline
\cline{3-6} \cline{4-6} \cline{5-6} \cline{6-6} 
 &  & {\small{}9992} & \cite{Seely:2009gt} & {\small{}26} & 16\tabularnewline
\cline{3-6} \cline{4-6} \cline{5-6} \cline{6-6} 
 &  & {\small{}9993} & \cite{Seely:2009gt} & {\small{}18} & 6\tabularnewline
\cline{3-6} \cline{4-6} \cline{5-6} \cline{6-6} 
 &  & {\small{}9994} & \cite{Seely:2009gt} & {\small{}17} & 7\tabularnewline
\cline{3-6} \cline{4-6} \cline{5-6} \cline{6-6} 
 &  & {\small{}9995} & \cite{Seely:2009gt} & {\small{}15} & 2\tabularnewline
\cline{3-6} \cline{4-6} \cline{5-6} \cline{6-6} 
 &  & {\small{}9996} & \cite{Seely:2009gt} & {\small{}19} & 7\tabularnewline
\cline{3-6} \cline{4-6} \cline{5-6} \cline{6-6} 
 &  & {\small{}9997} & \cite{Seely:2009gt} & {\small{}16} & 2\tabularnewline
\cline{3-6} \cline{4-6} \cline{5-6} \cline{6-6} 
 &  & {\small{}9998} & \cite{Seely:2009gt} & {\small{}21} & 8\tabularnewline
\cline{3-6} \cline{4-6} \cline{5-6} \cline{6-6} 
 &  & {\small{}9999} & \cite{Seely:2009gt} & {\small{}18} & 3\tabularnewline
\hline 
\hline 
\textbf{\small{}Total} &  &  &  & {\small{}546} & 336\tabularnewline
\hline 
\end{tabular}
\caption{This table summarizes the new JLab data  used in the fit 
including the CLAS data from Ref.~\cite{Schmookler:2019nvf} and the Hall~C data from Ref.~\cite{Seely:2009gt}.
The last column shows the number of data points remaining after the  \ncteqhix{} cuts $Q{>}1.3$~GeV and $W{>}1.7$~GeV are applied.  
}  %
\label{tab:jlab}
\end{table}
\begin{table}[tbp]
\centering{}{\footnotesize{}}%
\begin{tabular}{|l|l|c|c|c|c|}
\hline 
{\footnotesize{}$\mathbf{F_{2}^{A}/F_{2}^{D}:}$} &  &  &  &  & {\footnotesize{}$\#$ data}\tabularnewline
{\footnotesize{}Observable} & {\footnotesize{}Experiment} & {\footnotesize{}ID} & {\footnotesize{}Ref.} & {\footnotesize{}$\#$ data} & {\footnotesize{}after cuts}\tabularnewline
\hline 
\hline 
{\footnotesize{}D} & {\footnotesize{}NMC-97} & {\footnotesize{}5160} & {\footnotesize{}\cite{Arneodo:1996qe}} & {\footnotesize{}292} & {\footnotesize{}201/275}\tabularnewline
\hline \hline
{\footnotesize{}${}^3$He/D} & {\footnotesize{}Hermes} & {\footnotesize{}5156} & {\footnotesize{}\cite{Airapetian:2002fx}} & {\footnotesize{}182} & {\footnotesize{}17/92}\tabularnewline
\hline
\multirow{2}{*}{{\footnotesize${}^4${}He/D}} & {\footnotesize{}NMC-95,re} & {\footnotesize{}5124} & {\footnotesize{}\cite{Amaudruz:1995tq}} & {\footnotesize{}18} & {\footnotesize{}12/16}\tabularnewline
\cline{2-6} \cline{3-6} \cline{4-6} \cline{5-6} \cline{6-6} 
 & {\footnotesize{}SLAC-E139} & {\footnotesize{}5141} & {\footnotesize{}\cite{Gomez:1993ri}} & {\footnotesize{}18} & {\footnotesize{}3/17}\tabularnewline
\hline 
{\footnotesize{}Li/D} & {\footnotesize{}NMC-95} & {\footnotesize{}5115} & {\footnotesize{}\cite{Arneodo:1995cs}} & {\footnotesize{}24} & {\footnotesize{}11/15}\tabularnewline
\hline 
{\footnotesize{}Be/D} & {\footnotesize{}SLAC-E139} & {\footnotesize{}5138} & {\footnotesize{}\cite{Gomez:1993ri}} & {\footnotesize{}17} & {\footnotesize{}3/16}\tabularnewline
\hline 
\multirow{6}{*}{{\footnotesize{}C/D}} & {\footnotesize{}FNAL-E665-95} & {\footnotesize{}5125} & {\footnotesize{}\cite{Adams:1995is}} & {\footnotesize{}11} & {\footnotesize{}3/4}\tabularnewline
\cline{2-6} \cline{3-6} \cline{4-6} \cline{5-6} \cline{6-6} 
 & {\footnotesize{}SLAC-E139} & {\footnotesize{}5139} & {\footnotesize{}\cite{Gomez:1993ri}} & {\footnotesize{}7} & {\footnotesize{}2/7}\tabularnewline
\cline{2-6} \cline{3-6} \cline{4-6} \cline{5-6} \cline{6-6} 
 & {\footnotesize{}EMC-88} & {\footnotesize{}5107} & {\footnotesize{}\cite{Ashman:1988bf}} & {\footnotesize{}9} & {\footnotesize{}9/9}\tabularnewline
\cline{2-6} \cline{3-6} \cline{4-6} \cline{5-6} \cline{6-6} 
 & {\footnotesize{}EMC-90} & {\footnotesize{}5110} & {\footnotesize{}\cite{Arneodo:1989sy}} & {\footnotesize{}9} & {\footnotesize{}0/2}\tabularnewline
\cline{2-6} \cline{3-6} \cline{4-6} \cline{5-6} \cline{6-6} 
 & {\footnotesize{}NMC-95} & {\footnotesize{}5113} & {\footnotesize{}\cite{Arneodo:1995cs}} & {\footnotesize{}24} & {\footnotesize{}12/15}\tabularnewline
\cline{2-6} \cline{3-6} \cline{4-6} \cline{5-6} \cline{6-6} 
 & {\footnotesize{}NMC-95,re} & {\footnotesize{}5114} & {\footnotesize{}\cite{Amaudruz:1995tq}} & {\footnotesize{}18} & {\footnotesize{}12/16}\tabularnewline
\hline 
\multirow{2}{*}{{\footnotesize{}N/D}} & {\footnotesize{}Hermes} & {\footnotesize{}5157} & {\footnotesize{}\cite{Airapetian:2002fx}} & {\footnotesize{}175} & {\footnotesize{}19/92}\tabularnewline
\cline{2-6} \cline{3-6} \cline{4-6} \cline{5-6} \cline{6-6} 
 & {\footnotesize{}BCDMS-85} & {\footnotesize{}5103} & {\footnotesize{}\cite{Bari:1985ga}} & {\footnotesize{}9} & {\footnotesize{}9/9}\tabularnewline
\hline 
\multirow{2}{*}{{\footnotesize{}Al/D}} & {\footnotesize{}SLAC-E049} & {\footnotesize{}5134} & {\footnotesize{}\cite{Bodek:1983ec}} & {\footnotesize{}18} & {\footnotesize{}0/18}\tabularnewline
\cline{2-6} \cline{3-6} \cline{4-6} \cline{5-6} \cline{6-6} 
 & {\footnotesize{}SLAC-E139} & {\footnotesize{}5136} & {\footnotesize{}\cite{Gomez:1993ri}} & {\footnotesize{}17} & {\footnotesize{}3/16}\tabularnewline
\hline 
\multirow{4}{*}{{\footnotesize{}Ca/D}} & {\footnotesize{}NMC-95,re} & {\footnotesize{}5121} & {\footnotesize{}\cite{Amaudruz:1995tq}} & {\footnotesize{}18} & {\footnotesize{}12/15}\tabularnewline
\cline{2-6} \cline{3-6} \cline{4-6} \cline{5-6} \cline{6-6} 
 & {\footnotesize{}FNAL-E665-95} & {\footnotesize{}5126} & {\footnotesize{}\cite{Adams:1995is}} & {\footnotesize{}11} & {\footnotesize{}3/4}\tabularnewline
\cline{2-6} \cline{3-6} \cline{4-6} \cline{5-6} \cline{6-6} 
 & {\footnotesize{}SLAC-E139} & {\footnotesize{}5140} & {\footnotesize{}\cite{Gomez:1993ri}} & {\footnotesize{}7} & {\footnotesize{}2/7}\tabularnewline
\cline{2-6} \cline{3-6} \cline{4-6} \cline{5-6} \cline{6-6} 
 & {\footnotesize{}EMC-90} & {\footnotesize{}5109} & {\footnotesize{}\cite{Arneodo:1989sy}} & {\footnotesize{}9} & {\footnotesize{}0/2}\tabularnewline
\hline 
\multirow{5}{*}{{\footnotesize{}Fe/D}} & {\footnotesize{}SLAC-E049} & {\footnotesize{}5131} & {\footnotesize{}\cite{Bodek:1983qn}} & {\footnotesize{}14} & {\footnotesize{}2/14}\tabularnewline
\cline{2-6} \cline{3-6} \cline{4-6} \cline{5-6} \cline{6-6} 
 & {\footnotesize{}SLAC-E139} & {\footnotesize{}5132} & {\footnotesize{}\cite{Gomez:1993ri}} & {\footnotesize{}23} & {\footnotesize{}6/22}\tabularnewline
\cline{2-6} \cline{3-6} \cline{4-6} \cline{5-6} \cline{6-6} 
 & {\footnotesize{}SLAC-E140} & {\footnotesize{}5133} & {\footnotesize{}\cite{Dasu:1993vk}} & {\footnotesize{}10} & {\footnotesize{}0/6}\tabularnewline
\cline{2-6} \cline{3-6} \cline{4-6} \cline{5-6} \cline{6-6} 
 & {\footnotesize{}BCDMS-87} & {\footnotesize{}5101} & {\footnotesize{}\cite{Benvenuti:1987az}} & {\footnotesize{}10} & {\footnotesize{}10/10}\tabularnewline
\cline{2-6} \cline{3-6} \cline{4-6} \cline{5-6} \cline{6-6} 
 & {\footnotesize{}BCDMS-85} & {\footnotesize{}5102} & {\footnotesize{}\cite{Bari:1985ga}} & {\footnotesize{}6} & {\footnotesize{}6/6}\tabularnewline
\hline 
\multirow{3}{*}{{\footnotesize{}Cu/D}} & {\footnotesize{}EMC-93} & {\footnotesize{}5104} & {\footnotesize{}\cite{Ashman:1992kv}} & {\footnotesize{}10} & {\footnotesize{}9/10}\tabularnewline
\cline{2-6} \cline{3-6} \cline{4-6} \cline{5-6} \cline{6-6} 
 & {\footnotesize{}EMC-93(chariot)} & {\footnotesize{}5105} & {\footnotesize{}\cite{Ashman:1992kv}} & {\footnotesize{}9} & {\footnotesize{}9/9}\tabularnewline
\cline{2-6} \cline{3-6} \cline{4-6} \cline{5-6} \cline{6-6} 
 & {\footnotesize{}EMC-88} & {\footnotesize{}5106} & {\footnotesize{}\cite{Ashman:1988bf}} & {\footnotesize{}9} & {\footnotesize{}9/9}\tabularnewline
\hline 
{\footnotesize{}Kr/D} & {\footnotesize{}Hermes} & {\footnotesize{}5158} & {\footnotesize{}\cite{Airapetian:2002fx}} & {\footnotesize{}167} & {\footnotesize{}12/84}\tabularnewline
\hline 
{\footnotesize{}Ag/D} & {\footnotesize{}SLAC-E139} & {\footnotesize{}5135} & {\footnotesize{}\cite{Gomez:1993ri}} & {\footnotesize{}7} & {\footnotesize{}2/7}\tabularnewline
\hline 
{\footnotesize{}Sn/D} & {\footnotesize{}EMC-88} & {\footnotesize{}5108} & {\footnotesize{}\cite{Ashman:1988bf}} & {\footnotesize{}8} & {\footnotesize{}8}/8\tabularnewline
\hline 
{\footnotesize{}Xe/D} & {\footnotesize{}FNAL-E665-92} & {\footnotesize{}5127} & {\footnotesize{}\cite{Adams:1992nf}} & {\footnotesize{}10} & {\footnotesize{}2/4}\tabularnewline
\hline 
{\footnotesize{}Au/D} & {\footnotesize{}SLAC-E139} & {\footnotesize{}5137} & {\footnotesize{}\cite{Gomez:1993ri}} & {\footnotesize{}18} & {\footnotesize{}3/17}\tabularnewline
\hline 
{\footnotesize{}Pb/D} & {\footnotesize{}FNAL-E665-95} & {\footnotesize{}5129} & {\footnotesize{}\cite{Adams:1995is}} & {\footnotesize{}11} & {\footnotesize{}3/4}\tabularnewline
\hline 
\hline 
\textbf{\footnotesize{}Total:} &  &  &  & {\footnotesize{}1205} & {\footnotesize{}414/857}\tabularnewline
\hline 
\end{tabular}\caption{The DIS $F_{2}^{A}/F_{2}^{D}$ data sets used in this fit. 
The table details  
the specific nuclear targets, the references, and the total number of data points
in the set. 
The last column shows first the number of data points remaining 
with the nCTEQ15 cuts of  $Q{>}2$~GeV and $W{>}3.5$~GeV,
and second   the number of data points remaining 
with the \ncteqhix{}  cuts of  $Q{>}1.3$~GeV and $W{>}1.7$~GeV.
The NMC data set ID=5160 is $F_2^D$; this is the one case which is not a ratio.
}%
\label{tab:exp1}
\end{table}
\begin{table}[tpb]
\centering{}{\footnotesize{}}%
\begin{tabular}{|l|l|c|c|c|c|}
\hline 
{\footnotesize{}$\mathbf{F_{2}^{A}/F_{2}^{A'}:}$} &  &  &  &  & {\footnotesize{}$\#$ data}\tabularnewline
{\footnotesize{}Observable} & {\footnotesize{}Experiment} & {\footnotesize{}ID} & {\footnotesize{}Ref.} & {\footnotesize{}$\#$ data} & {\footnotesize{}after cuts}\tabularnewline
\hline 
\hline 
{\footnotesize{}C/Li} & {\footnotesize{}NMC-95,re} & {\footnotesize{}5123} & {\footnotesize{}\cite{Amaudruz:1995tq}} & {\footnotesize{}25} & {\footnotesize{}7/20}\tabularnewline
\hline 
{\footnotesize{}Ca/Li} & {\footnotesize{}NMC-95,re} & {\footnotesize{}5122} & {\footnotesize{}\cite{Amaudruz:1995tq}} & {\footnotesize{}25} & {\footnotesize{}7/20}\tabularnewline
\hline 
{\footnotesize{}Be/C} & {\footnotesize{}NMC-96} & {\footnotesize{}5112} & {\footnotesize{}\cite{Arneodo:1996rv}} & {\footnotesize{}15} & {\footnotesize{}14/15}\tabularnewline
\hline 
{\footnotesize{}Al/C} & {\footnotesize{}NMC-96} & {\footnotesize{}5111} & {\footnotesize{}\cite{Arneodo:1996rv}} & {\footnotesize{}15} & {\footnotesize{}14/15}\tabularnewline
\hline 
\multirow{2}{*}{{\footnotesize{}Ca/C}} & {\footnotesize{}NMC-95,re} & {\footnotesize{}5120} & {\footnotesize{}\cite{Amaudruz:1995tq}} & {\footnotesize{}25} & {\footnotesize{}7/20}\tabularnewline
\cline{2-6} \cline{3-6} \cline{4-6} \cline{5-6} \cline{6-6} 
 & {\footnotesize{}NMC-96} & {\footnotesize{}5119} & {\footnotesize{}\cite{Arneodo:1996rv}} & {\footnotesize{}15} & {\footnotesize{}14/15}\tabularnewline
\hline 
{\footnotesize{}Fe/C} & {\footnotesize{}NMC-96} & {\footnotesize{}5143} & {\footnotesize{}\cite{Arneodo:1996rv}} & {\footnotesize{}15} & {\footnotesize{}14/15}\tabularnewline
\hline 
{\footnotesize{}Sn/C} & {\footnotesize{}NMC-96} & {\footnotesize{}5159} & {\footnotesize{}\cite{Arneodo:1996ru}} & {\footnotesize{}146} & {\footnotesize{}111/144}\tabularnewline
\hline 
{\footnotesize{}Pb/C} & {\footnotesize{}NMC-96} & {\footnotesize{}5116} & {\footnotesize{}\cite{Arneodo:1996rv}} & {\footnotesize{}15} & {\footnotesize{}14/15}\tabularnewline
\hline 
\hline 
\textbf{\footnotesize{}Total:} &  &  &  & {\footnotesize{}296} & {\footnotesize{}202/279}\tabularnewline
\hline 
\end{tabular}\caption{The DIS $F_{2}^{A}/F_{2}^{A'}$ data sets used in this
fit. We list the same details for each data set as in Tab.~\ref{tab:exp1}.}
\label{tab:exp2}
\end{table}
\begin{table}[tbp]
\centering{}{\footnotesize{}}%
\begin{tabular}{|l|l|c|c|c|c|}
\hline 
{\footnotesize{}$\mathbf{\sigma_{DY}^{pA}/\sigma_{DY}^{pA'}:}$} &  &  &  &  & {\footnotesize{}$\#$ data}\tabularnewline
{\footnotesize{}Observable} & {\footnotesize{}Experiment} & {\footnotesize{}ID} & {\footnotesize{}Ref.} & {\footnotesize{}$\#$ data} & {\footnotesize{}after cuts}\tabularnewline
\hline 
\hline 
{\footnotesize{}C/D} & {\footnotesize{}FNAL-E772-90} & {\footnotesize{}5203} & {\footnotesize{}\cite{Alde:1990im}} & {\footnotesize{}9} & {\footnotesize{}9/9}\tabularnewline
\hline 
{\footnotesize{}Ca/D} & {\footnotesize{}FNAL-E772-90} & {\footnotesize{}5204} & {\footnotesize{}\cite{Alde:1990im}} & {\footnotesize{}9} & {\footnotesize{}9/9}\tabularnewline
\hline 
{\footnotesize{}Fe/D} & {\footnotesize{}FNAL-E772-90} & {\footnotesize{}5205} & {\footnotesize{}\cite{Alde:1990im}} & {\footnotesize{}9} & {\footnotesize{}9/9}\tabularnewline
\hline 
{\footnotesize{}W/D} & {\footnotesize{}FNAL-E772-90} & {\footnotesize{}5206} & {\footnotesize{}\cite{Alde:1990im}} & {\footnotesize{}9} & {\footnotesize{}9/9}\tabularnewline
\hline 
{\footnotesize{}Fe/Be} & {\footnotesize{}FNAL-E886-99} & {\footnotesize{}5201} & {\footnotesize{}\cite{Vasilev:1999fa}} & {\footnotesize{}28} & {\footnotesize{}28/28}\tabularnewline
\hline 
{\footnotesize{}W/Be} & {\footnotesize{}FNAL-E886-99} & {\footnotesize{}5202} & {\footnotesize{}\cite{Vasilev:1999fa}} & {\footnotesize{}28} & {\footnotesize{}28/28}\tabularnewline
\hline 
\hline 
\textbf{\footnotesize{}Total:} &  &  &  & {\footnotesize{}92} & {\footnotesize{}92/92}\tabularnewline
\hline 
\end{tabular}\caption{{ The Drell-Yan process data sets used in this fit.
We list the same details for each data set as in Tab.~\ref{tab:exp1}.
}}
\label{tab:exp3}
\end{table}
\bibliographystyle{utphys}

\cleardoublepage
\bibliography{extra,main}

\clearpage{}

\begin{widetext}
\printfigures
\printtables 
\end{widetext}

\end{document}